\documentclass[twocolumn]{aastex7}

\usepackage{gensymb}
\newcommand{\src}{G21.5$-$0.9}
\newcommand{\chandra}{CXO}
\newcommand{\ixpeobssim}{\textsc{ixpeobssim}}
\newcommand{\ixpesim}{\textsc{ixpesim}}
\usepackage{wrapfig}

\begin{document}

\title{X-ray Polarization Detection of the Pulsar Wind Nebula in \src{} with IXPE}

\correspondingauthor{Niccol\`{o} Di Lalla}
\author[0000-0002-7574-1298]{Niccol\`{o} Di Lalla}
\affiliation{W. W. Hansen Experimental Physics Laboratory (HEPL), Stanford University, Stanford, California 94305, USA}
\affiliation{Department of Physics and Kavli Institute for Particle Astrophysics and Cosmology, Stanford University, Stanford, California 94305, USA}
\email[show]{niccolo.dilalla@stanford.edu}

\author[0000-0002-5448-7577]{Nicola Omodei}
\affiliation{W. W. Hansen Experimental Physics Laboratory (HEPL), Stanford University, Stanford, California 94305, USA}
\affiliation{Department of Physics and Kavli Institute for Particle Astrophysics and Cosmology, Stanford University, Stanford, California 94305, USA}
\email{nicola.omodei@stanford.edu}

\author[0000-0002-8848-1392]{Niccol\`{o} Bucciantini}
\affiliation{INAF Osservatorio Astrofisico di Arcetri, Largo Enrico Fermi 5, 50125 Firenze, Italy}
\affiliation{Dipartimento di Fisica e Astronomia, Universit\`{a} degli Studi di Firenze, Via Sansone 1, 50019 Sesto Fiorentino (FI), Italy}
\affiliation{Istituto Nazionale di Fisica Nucleare, Sezione di Firenze, Via Sansone 1, 50019 Sesto Fiorentino (FI), Italy}
\email{niccolo.bucciantini@inaf.it}

\author[0000-0002-6401-778X]{Jack T. Dinsmore}
\affiliation{Department of Physics and Kavli Institute for Particle Astrophysics and Cosmology, Stanford University, Stanford, California 94305, USA}
\email{jtd@stanford.edu}

\author[0000-0003-3842-4493]{Nicol\`{o} Cibrario}
\affiliation{Istituto Nazionale di Fisica Nucleare, Sezione di Torino, Via Pietro Giuria 1, 10125 Torino, Italy}
\affiliation{Dipartimento di Fisica, Universit\`{a} degli Studi di Torino, Via Pietro Giuria 1, 10125 Torino, Italy}
\email{nicolo.cibrario@unito.it}

\author[0000-0002-8665-0105]{Stefano Silvestri}
\affiliation{Istituto Nazionale di Fisica Nucleare, Sezione di Pisa, Largo B. Pontecorvo 3, 56127 Pisa, Italy}
\email{stefano.silvestri@pi.infn.it}

\author[0000-0001-6395-2066]{Josephine Wong}
\affiliation{Department of Physics and Kavli Institute for Particle Astrophysics and Cosmology, Stanford University, Stanford, California 94305, USA}
\email{joswong@stanford.edu}

\author[0000-0002-6986-6756]{Patrick Slane}
\affiliation{Center for Astrophysics | Harvard \& Smithsonian, 60 Garden Street, Cambridge, MA 02138, USA}
\email{pslane@cfa.harvard.edu}

\author[0000-0001-7263-0296]{Tsunefumi Mizuno}
\affiliation{Hiroshima Astrophysical Science Center, Hiroshima University, 1-3-1 Kagamiyama, Higashi-Hiroshima, Hiroshima 739-8526, Japan}
\email{mizuno@astro.hiroshima-u.ac.jp}

\author[0000-0002-6548-5622]{Michela Negro}
\affiliation{Department of Physics and Astronomy, Louisiana State University, Baton Rouge, LA 70803 USA}
\email{michelanegro@lsu.edu}

% ...


\begin{abstract}
We present the X-ray polarization observation of \src{}, a young Galactic supernova remnant (SNR), conducted with the Imaging X-ray Polarimetry Explorer (IXPE) in October 2023, with a total livetime of approximately 837~ks. Using different analysis methods, such as a space-integrated study of the entire region of the PWN and a space-resolved polarization map, we detect significant polarization from the pulsar wind nebula (PWN) at the center of the SNR, with an average polarization degree of $\sim 10\%$ oriented at $\sim 33\degree$ (north through east). No significant energy-dependent variation in polarization is observed across the IXPE band (2--8~keV). The polarization map, corrected for the effect of polarization leakage, reveals a consistent pattern in both degree and angle, with little change across the nebula. Our findings indicate the presence of a highly polarized central torus, suggesting low levels of turbulence at particle acceleration sites. Unlike Vela, but similar to the Crab Nebula, we observe substantial differences between radio and X-ray polarization maps. This suggests a clear separation in energy of the emitting particle populations and hints at an important, yet poorly understood, role of instabilities in the turbulence dynamics of PWNe.
\end{abstract}



\section{Introduction} \label{sec:int}
Pulsars are strongly magnetized and rapidly rotating neutron stars, born from the explosive deaths of massive stars, first discovered by Jocelyn Bell Burnell in 1967 using the radio telescope, Interplanetary Scintillation Array in Cambridge, UK~\citep{1968Natur.217..709H}. As pulsars spin around their axes, they unleash powerful beams of radiation and a persistent powerful stream of relativistic particles into space. These emitted particles, which together with the Poynting flux form the so-called \textit{pulsar wind}, are accelerated to nearly the speed of light by the intense magnetic and electric fields present in the magnetosphere.
As they propagate along the magnetic field lines, they interact with the surrounding interstellar medium or supernova ejecta. This interaction creates shock fronts and gives rise to an extended and complex synchrotron-emitting nebula, from radio to X-ray and MeV energies, called pulsar wind nebula (PWN)~\citep{Gaensler2006}. PWNe constitute an incredibly fascinating laboratory for exploring fundamental physics in conditions not replicable on Earth, providing valuable insights into relativistic shocks and particle acceleration processes governing the dynamics of relativistic outflows and their influence on the surrounding interstellar environment.

\src{} is a young, composite \textit{plerionic} supernova remnant (SNR) characterized in X-rays by the presence of a central, bright, 40\arcsec\ radius PWN surrounded by a much dimmer and softer SNR shell~\citep{2000ApJ...533L..29S, 2001ApJ...561..308S}. Deep observations obtained with \textit{Chandra} X-ray Observatory (\chandra{}) and XMM-Newton have revealed that the SNR blast wave extends up to 150\arcsec\ and displays a limb brightening feature at the eastern boundary and knots of enhanced soft X-ray emission above the PWN in the northern direction~\citep{MATHESON20051099,2005A&A...442..539B,2010ApJ...724..572M,2019MNRAS.482.1031G}.
Observations of \src{} across various electromagnetic wavelengths, from radio to $\gamma$-rays, have unveiled a wealth of information about this source. Using the Very Large Array (VLA) to measure its expansion speed, \citet{2008MNRAS.386.1411B} estimated its age to be $\sim 870$~yr, making \src{} one of the youngest PWNe known in our galaxy. Its distance to Earth was measured by several authors using different techniques and was found to be around 5~kpc \citep{2001ApJ...561..308S,2006ApJ...637..456C,2008MNRAS.386.1411B, 2008MNRAS.391L..54T}. The first maps at radio wavelengths go back to the 1970--80s~\citep{1976A&A....53...89W,1976ApJ...204..427B,1988PASJ...40..347F}, but even more recent deep searches have failed to detect a radio counterpart of the X-ray SNR shell, with the exception of the northern knot~\citep{2011MNRAS.412.1221B}.

At the heart of \src{} lies the pulsar PSR J1833$-$1034, first independently discovered in radio by \citet{2005CSci...89..853G} and \citet{2006ApJ...637..456C}, who found a compact pulsating source at the center of the PWN (R.A. = $18^h33^m33^s.57$, Dec. = $-10\degree34'07''.5$ (J2000.0)) with period $P = 61.86$~ms, $\dot P = 2\times10^{-13}$, surface magnetic field of $B = 3.6 \times 10^{12}$~G, characteristic age of 4.8~kyr, and a spin-down luminosity of $\dot E = 3.3 \times 10^{37} \; \mathrm{erg \: s^{-1}}$. The pulsar remains currently undetected at X-ray energies, while $\gamma$-ray pulsation has been detected with the \textit{Fermi} Large Area Telescope~\citep{2013ApJS..208...17A, 2023ApJ...958..191S}. Past X-ray observations have only led to upper limits on the pulsating fraction, while still detecting a central compact source surrounded by an elliptical emission region~\citep{2000ApJ...533L..29S, 2001ApJ...561..308S, 2006ApJ...637..456C}. The projected spin-axis direction of PSR J1833$-$1034, inferred by fitting the pulsar wind torus using \chandra{} data, was determined to be at a position angle of $45\degree \pm 1\degree$ north through east~\citep{2008ApJ...673..411N}.

Polarization observations can play a crucial role in unraveling the complex structure of PWNe like \src{}, by offering unique insights into the physical processes at play within these astrophysical sources. Polarimetry, in fact, provides essential information about magnetic fields, particle acceleration mechanisms, and the overall dynamics of these systems, contributing to a more comprehensive understanding of the intricate interplay between pulsars and their cosmic surroundings.

The initial radio polarization maps of \src{} were obtained by \citet{1981ApJ...248L..23B} with the VLA and by \citet{1988PASJ...40..347F} using the NRO Millimeter-wave Array and revealed a highly ordered, linearly polarized emission distributed in a ring structure, indicating a radial magnetic field configuration. The level of polarization was found to have a minimum towards the center, where little polarization was detected, and to increase with radius up to 20--30\% near the boundary of the PWN. Infrared polarimetric observations performed with the Very Large Telescope on the compact core of the PWN (angular size $\lesssim~$5\arcsec) revealed a highly polarized emission, with an average value of the linear polarization degree of $\sim 47\%$, and a swing of the polarization angle across the inner nebula consistent with a toroidal magnetic field~\citep{2012A&A...542A..12Z}. On the other hand, a recent polarimetric study by \citet{2022ApJ...930....1L}, using archival high-resolution radio observations taken with VLA, confirmed the global radial magnetic field structure of the PWN and ruled out the existence of any strong large-scale toroidal field components extending beyond the inner nebula.

Unlike Crab, Vela, or MSH 15$-$52 \citep{Bucciantini23a,Xie22,Romani23a}, which in X-rays all display a morphology dominated by a jet-torus structure, \src{} has a much more amorphous appearance, the torus is barely evident, and there is no sign of a jet. Moreover, among the PWNe that can be resolved by IXPE, \src{} is the only one that shows a tangential polarization pattern in radio. In this regard, this source represents a system quite different from the ones studied previously, and X-ray polarimetry could provide insight to better understand such a difference.

This paper is structured as follows. In Sect.~\ref{sec:obs} we present the IXPE observation of \src{}, together with the initial processing done on the data. Sect.~\ref{sec:pol} is dedicated to the description of the different types of polarization analyses performed and to the presentation of the main results. The implications and possible physical interpretation of these results are discussed in Sect.~\ref{sec:dis}. Finally, Appendices~\ref{sec:app_3ml} and \ref{sec:app_leak} contain, respectively, additional plots of the polarization analysis and a review of the methods used to evaluate or subtract the effect of polarization leakage.

\section{Observations} \label{sec:obs}
The Imaging X-ray Polarimetry Explorer (IXPE), successfully launched on December 9, 2021, is the first space observatory entirely dedicated to imaging polarimetry in X-rays~\citep{Weisskopf2022}. Result of an international collaboration between NASA and the Italian Space Agency (ASI), IXPE is provided with three identical but independent telescopes, each equipped with a Gas Pixel Detector (GPD) at its focal plane~\citep{2006NIMPA.560..425B, 2021APh...13302628B}, housed inside a detector unit (DU). 

\begin{figure}[t]
    \centering
    \includegraphics[trim={0 0 1.5cm 0.8cm},clip,width=0.5\textwidth]{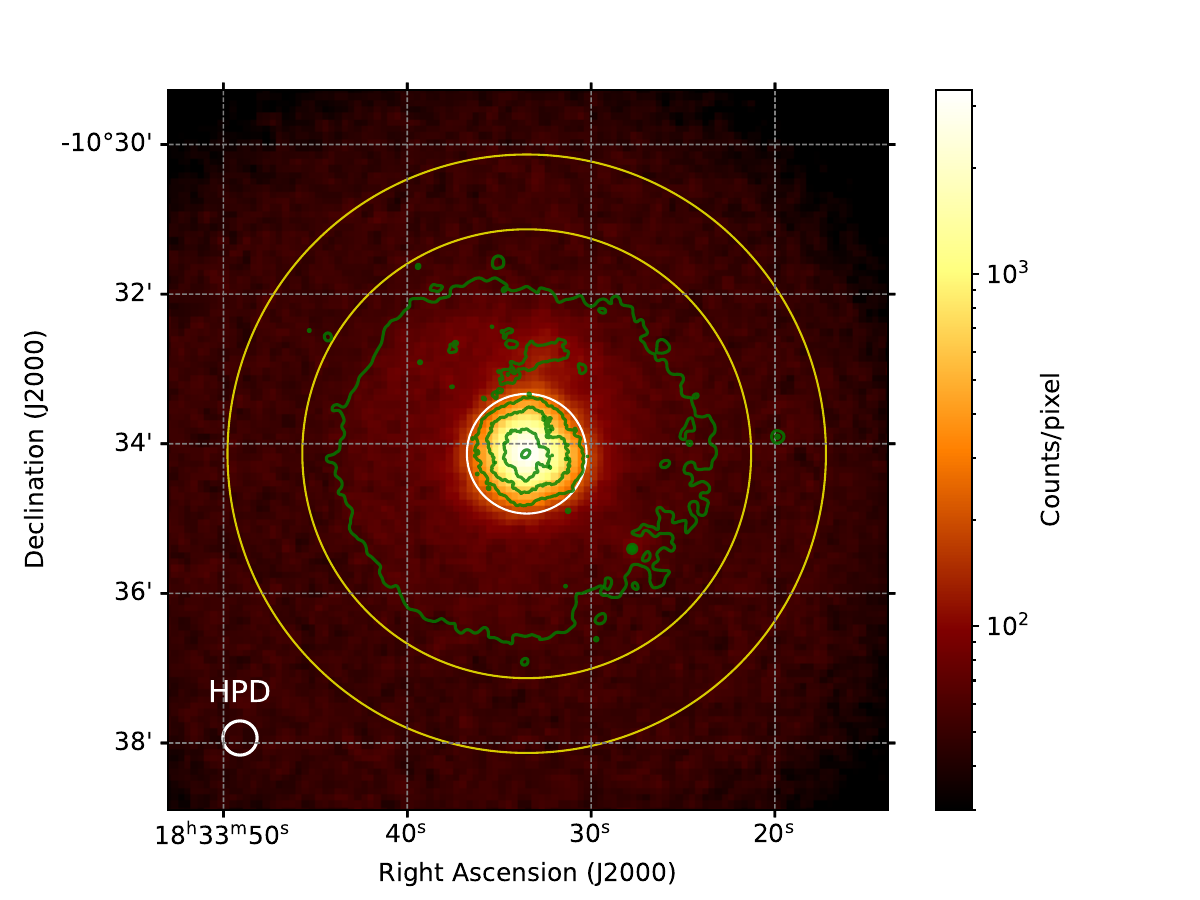}
    \caption{IXPE smoothed count map of \src{} in logarithmic scale, obtained by combining the data from the three DUs (2--8~keV). In green are the contours from the deep \chandra{} merged observations (0.5--8~keV), described in the text and visible as a background image of Fig.~\ref{fig:pmap}. The white circle and yellow annulus indicate respectively the source and background extraction regions used for the polarization analysis described in Sect.~\ref{sec:spint}. At the bottom left corner is the IXPE half-power diameter HPD~$\sim$~25\arcsec\ for DU1 for reference.}
    \label{fig:cmap}
\end{figure}

IXPE observed \src{} in three separate segments close in time and combined in a single observation with a total livetime of $\sim 837$~ks: the first segment lasted from October 5, 2023 at 00:05:39.233 UTC to October 9, 2023 at 02:16:33.292 UTC, the second one from October 10, 2023 at 22:45:54.503 UTC to October 17, 2023 at 16:42:55.034 UTC and the last one from October 19, 2023 at 21:35:44.584 UTC to October 27, 2023 at 15:39:00.534 UTC. IXPE Level-2 data, publicly available for download on the HEASARC archive\footnote{\url{https://heasarc.gsfc.nasa.gov/docs/ixpe/archive/}} (ObsID 02001199), were reprocessed and analyzed according to the standard technique using the version \texttt{20240125} of the IXPE Instrument Response Functions (IRFs) released on the CALDB\footnote{\url{https://heasarc.gsfc.nasa.gov/docs/ixpe/caldb/}} on February 28, 2024 (validity date \texttt{20230702}). Given the temporal proximity and relatively short total duration of the IXPE observations, spanning less than a month, and the known variability timescale of the nebula, which is on the order of several months to years~\citep{2019MNRAS.482.1031G}, \src{} can be considered effectively steady for the purpose of this analysis. 

Before analyzing the data, we performed an initial background rejection following the procedure described in \citet{2023AJ....165..143D} to reduce the instrumental background due to cosmic-ray particles interacting in the GPD by approximately 40\%. We also excluded approximately 4~ks of total livetime ($\sim 0.5\%$) affected by sudden spikes in the counting rate over short time intervals, likely associated with solar activity and the boundary of the South Atlantic Anomaly. In addition, we corrected the absolute sky coordinates by adjusting the WCS keywords in the FITS header to align the reference frame with that of \chandra{} (\texttt{TCRPX7 = 304}, \texttt{TCRPX8 = 301.5}). These are the data files used from now on in this paper. 
Fig.~\ref{fig:cmap} shows the count map resulting from the combination of the three IXPE telescopes over the full energy range (2--8~keV). The white circular region (radius 0.8\arcmin\ centered on the pulsar location) is used to select the inner PWN, while the yellow annulus (inner/outer radii of 3\arcmin\ and 4\arcmin\ from the pulsar) defines the background extraction region used for the space-integrated analysis described in Sect.~\ref{sec:spint}.

\src{} is observed with \chandra{} routinely as a calibration source. Overlaid in green in Fig.~\ref{fig:cmap} are the contours of a deep image created by merging 18 individual observations listed in Table~\ref{tab:obsid} (total exposure $\sim 183$~ks) and contained in the \textit{Chandra} Data Collection~\dataset[DOI:10.25574/cdc.425]{https://doi.org/10.25574/cdc.425}, all of which had the remnant placed in the ACIS-S detector, at the default aimpoint of the S3 CCD chip. The events were combined with the \texttt{merge\_obs} script in the \texttt{ciao}\footnote{\url{https://cxc.cfa.harvard.edu/ciao/}} analysis software. A count map was then created using events in the 0.5--8.0~keV energy band, and adaptively smoothed with the \texttt{dmimgadapt} tool using a cone filter with minimum/maximum smoothing radii of 0.5\arcsec\ and 150\arcsec\ spanning 300 logarithmically spaced scales, with a minimum of 75 counts within the smoothing kernel. The resulting map is used as background image in all polarization maps shown throughout the paper.

\begin{deluxetable}{lclc}
\label{tab:obsid}
\tabletypesize{\scriptsize}
\tablewidth{0pt}
\tablecaption{List of the \chandra{} ObsIDs used.}
\tablehead{
\colhead{ObsID} & \colhead{Exposure (ks)} & \colhead{ObsID} & \colhead{Exposure (ks)}}
    \startdata 
    159 & 14.85 & 3700 &  9.54 \\
    1230 & 14.56 & 5159 &  9.83 \\
    1433 & 14.97 & 5166 & 10.02 \\
    1554 &  9.06 & 6071 &  9.64 \\
    1717 &  7.54 & 6741 &  9.83 \\
    1770 &  7.22 & 8372 & 10.01 \\
    1838 &  7.85 & 10646 &  9.54 \\
    2873 &  9.83 & 14263 &  9.57 \\
    3693 &  9.78 & 16420 &  9.57
    \enddata
    \tablecomments{List of ObsIDs and their exposure for the \chandra{} observations used to produce the count map used as background in all polarization maps shown in the paper (e.g. Fig.~\ref{fig:pmap}) and whose contours are shown in Fig.~\ref{fig:cmap}. The total exposure is around 183~ks. The \chandra{} datasets are contained in the \textit{Chandra} Data Collection~\dataset[DOI:10.25574/cdc.425]{https://doi.org/10.25574/cdc.425}.}
\end{deluxetable}

\section{Polarization Analysis} \label{sec:pol}
In this section, we describe the different analysis methods used to measure the polarization of \src. We performed two main types of analysis: an integrated study of the entire region of the PWN (Sect.~\ref{sec:spint}) and a space-resolved measurement aimed at mapping the polarization pattern across the nebula (Sect.~\ref{sec:spres}).

\subsection{Space-integrated analysis} \label{sec:spint}
\begin{figure}[b]
    \centering
    \includegraphics[trim={0.3cm 0.2cm 0cm 0cm},clip,width=0.42\textwidth]{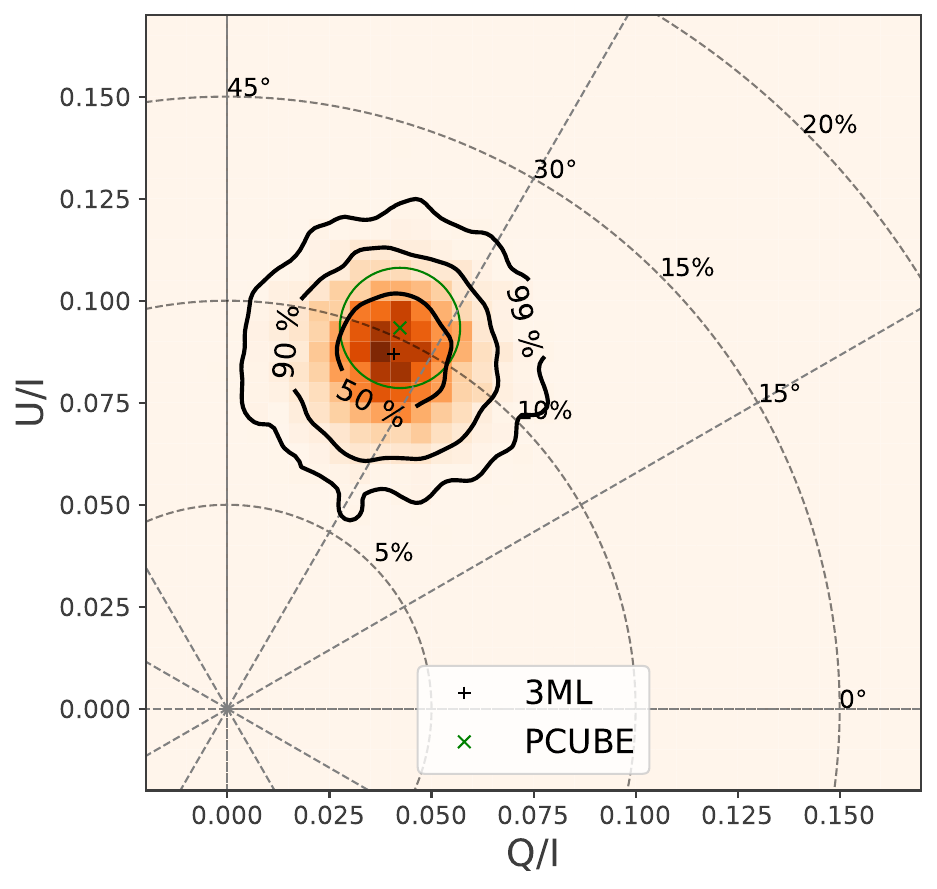}
    \caption{Q/I vs. U/I plot showing the results of the space-integrated polarization analysis of the PWN in the 2--8 keV energy range. In orange shades, the 2D distribution resulting from the spectro-polarimetric analysis with the $50\%$, $90\%$, and $99\%$ C.L. contours in black and the $+$ marker indicating the best-fit parameters. The green $\times$ and circle show respectively the result of the \texttt{PCUBE} analysis and the associated $1\sigma$ error.}
    \label{fig:pcube}
\end{figure}

As is typically done for other IXPE observations~\citep{Bucciantini23a,Xie22,2023ApJ...946L..21N}, using the spatial regions described at the end of Sect.~\ref{sec:obs} and shown in Fig.~\ref{fig:cmap}, we studied the polarization of the entire PWN in two different ways: a model-independent polarimetric analysis using \ixpeobssim~\citep{Baldini2022} and a spectro-polarimetric fit with the Multi-Mission Maximum Likelihood (\texttt{3ML}) framework~\citep{2015arXiv150708343V}. For the former, we utilized the \texttt{xpselect} and \texttt{xpbin} applications of \ixpeobssim~(v31.0.1) to first select photons belonging to the source and background regions and then bin these filtered data files using the \texttt{PCUBE} algorithm (unweighted analysis). This binning routine combines the event-by-event Stokes parameters, subtracts the background contribution, and computes the degree (PD) and angle (PA) of polarization of the source, along with the associated errors, according to the procedure of~\citet{Stokes}. PA represents the electric vector position angle and is conventionally defined anticlockwise relative to the local north (towards northeast). Combining the data from the three IXPE DUs over the 2--8~keV energy range, we measured with this method a PD of $10.2\% \pm 1.5\%$ for the PWN, with a PA of $33\degree \pm 4\degree$. The corresponding Minimum Detectable Polarization at 99\% confidence is $\mathrm{MDP}_{99\%} = 4.3\%$~\citep{Weisskopf_2010}, well below the detected level polarization. The measured PA is broadly consistent with the spin-axis direction of PSR J1833$-$1034, although in tension at 3$\sigma$ level. To investigate possible energy dependence, we repeated the analysis in two energy bands, 2--4~keV and 4--8~keV, and found no significant variation in polarization, with results consistent within uncertainties.  As an internal consistency check, we also repeated the analysis independently for each DU and found agreement within $1\sigma$ for both PD and PA. The results for the combined DUs are summarized on the left side of Table~\ref{tab:summary} and visualized in Fig.~\ref{fig:pcube}.

\begin{deluxetable*}{lcrlr}
\label{tab:summary}
\tabletypesize{\scriptsize}
\tablewidth{0pt}
\tablecaption{Summary of the space-integrated analysis of the PWN.}
\tablehead{
\colhead{Parameter} & \colhead{Energy range} & \colhead{Value} & \colhead{Parameter} & \colhead{Value}}
    \startdata 
            & 2--8~keV & $10.2\% \pm 1.5\%$ & $N_{\rm H}$ (fixed) & $3.237 \times 10^{22} \; \mathrm{cm}^{-2}$ \\
        PD  & 2--4~keV & $9.7\% \pm 1.5\%$ & $\Gamma$ & $1.92 \pm 0.02$ \\
            & 4--8~keV & $11.3\% \pm 2.5\%$ & PD & $9.7\% \pm 1.2\%$ \\ \cline{1-3}
            & 2--8~keV & $33\degree\pm4\degree$ & PA & $32\degree \pm 4\degree$ \\
        PA  & 2--4~keV & $36\degree\pm4\degree$ & Q/I & $0.041 \pm 0.012$ \\
            & 4--8~keV & $28\degree\pm7\degree$ & U/I & $0.087 \pm 0.012$ \\
    \enddata
    \tablecomments{Summary table of the \texttt{PCUBE} analysis (left) and spectro-polarimetric fit (right) of the space-integrated PWN. Errors are statistical only at 1$\sigma$ level. The spectro-polarimetric fit is performed using \texttt{3ML} in the 2--8 keV energy range assuming Gaussian statistics and resulted in a reduced chi-square value of $\chi^2/\text{d.o.f.} = 1462.6/1353$ ($p$-value = 0.02).}
\end{deluxetable*}

For the spectro-polarimetric analysis, instead, we started from the same filtered data, produced earlier for the source and background regions, and used the \texttt{xpbin} application (algorithms \texttt{PHA1}, \texttt{PHA1Q} and \texttt{PHA1U}, unweighted analysis) to compute the Stokes I, Q, and U spectra for the three DUs. These files can then be jointly fit in \texttt{3ML} or \textsc{xspec}\footnote{\url{https://heasarc.gsfc.nasa.gov/xanadu/xspec/}} to simultaneously recover the spectral and polarimetric parameters of the user-defined models. For \src, we modeled the observed spectrum with an absorbed power-law, with absorption given by the \texttt{TbAbs}\footnote{\url{https://astromodels.readthedocs.io/en/latest/notebooks/TbAbs.html}} model with \textit{wilm} abundances~\citep{2000ApJ...542..914W}. According to what was found by~\citet{2019MNRAS.482.1031G}, the column density parameter was fixed to the value $N_{\rm H} = 3.237 \times 10^{22} \; \mathrm{cm}^{-2}$. Regarding the polarimetric model, based on the result of the previous \texttt{PCUBE} analysis, we chose to employ a model with a constant degree and angle of polarization. To account for potential cross-calibration uncertainties among the different IXPE telescopes, we also added and left free to vary a multiplicative normalization constant for DU2 and DU3 relative to DU1. The best-fit results obtained using \texttt{3ML} are provided on the right side of Table~\ref{tab:summary}. The power-law index we measured is $\Gamma = 1.92 \pm 0.02$, slightly higher than the value reported by~\citet{2019MNRAS.482.1031G}, most likely due to contamination from the much softer external halo and differences in the source extraction region. In support of this interpretation, our spectral result is in good agreement with the analysis by~\citet{2011A&A...525A..25T}, which used a much larger extraction radius (165\arcsec) encompassing the entire remnant. Note that the quoted uncertainty represents the statistical error only and does not account for possible systematic effects from instrumental calibration (see Appendix~\ref{sec:app_3ml} for further discussion). The unabsorbed flux (2--8~keV) is $F_{\mathrm{2-8 keV}} = (3.91 \pm 0.03) \times 10^{-11} \, \mathrm{erg \, s^{-1} cm^{-2}}$. The polarimetric results agree well with those from the \texttt{PCUBE} analysis, with a best-fit PD = $9.7\% \pm 1.2\%$ and PA = $32\degree \pm 4\degree$. Fig.~\ref{fig:pcube} gives a summary of the main results of the space-integrated polarimetric analysis of the PWN. 

\subsection{Space-resolved analysis} \label{sec:spres}

To map the polarization of \src{} in the PWN region, we again used the \texttt{xpbin} tool of \ixpeobssim, this time with the \texttt{PMAP} algorithm. This routine processes the provided data files by binning the Stokes parameters I, Q and U in sky coordinates, thus allowing us to measure the polarization pattern across the field of view. In particular, we binned the data using a grid of $90 \times 90$ squared pixels, each with a side of 0.1944\arcmin, considering photons with energies in the full IXPE energy band (2--8~keV). Then, we used the implemented methods to perform a two-dimensional convolution of the original Stokes maps with a unit-filled $3 \times 3$ kernel matrix, assigning to each pixel the sum of the contents of itself and its eight neighbors. In this way, adjacent pixels become highly correlated and the effective size of the pixel triples and becomes comparable to the IXPE angular resolution (half-power diameter $\sim$~25--30\arcsec), while still maintaining the original binning grid.

Since the start of the IXPE mission, it has been well known that imperfections in reconstructing photon absorption points within the GPD, arising from the event reconstruction algorithm~\citep{2021APh...13302628B} or from the finite spatial resolution of the GPD, can produce a radially polarized halo around the source, commonly referred to as \textit{polarization leakage} (for a detailed overview of the effect, see \cite{Bucciantini2023Leakage}). Being an intrinsic effect of the detector or its reconstruction algorithm, this phenomenon is present in all IXPE observations, but in practice it is only really relevant for the analysis of a handful of bright extended sources, which exhibit a very sharp intensity gradient.
Since the polarization pattern induced by the leakage is radial, its contribution effectively averages out when selecting a circular region that includes the entire source, as done in Sect.~\ref{sec:spint}. For the space-resolved polarization analysis, instead, polarization leakage can strongly affect the results, and a dedicated analysis is required to estimate and subtract its contribution. 

Given the morphology of \src{}, and its small relative size comparable to the IXPE angular resolution, it is expected that the effect of polarization leakage will be particularly relevant in the outskirts of the PWN. Its contribution was evaluated using different techniques, including a full \ixpesim/\ixpeobssim~\citep{dilalla2019} simulation of the source using \texttt{GEANT4}~\citep{GEANT4:2002zbu}, the Mueller matrix formalism described in \citet{Bucciantini2023Leakage}, a hybrid machine learning and analytic track reconstruction method~\citep{2023A&A...674A.107C} and, finally, the recently released \texttt{LeakageLib} package~\citep{Dinsmore2024}. The latter tool, in particular, implements the most accurate model of the polarization leakage to date using in-flight calibrated 2D PSFs, and its routines allow the user to predict the leakage pattern based on a \chandra{} image and subtract it from the measured IXPE polarization map. A detailed summary of the different approaches can be found in Appendix~\ref{sec:app_leak}. Despite methodological differences, all four techniques yield consistent polarization structures within uncertainties, confirming the robustness of the analysis, particularly in the central, high-significance region of the PWN.

\begin{figure}
    \centering
    \includegraphics[trim={0.1cm 0 2.3cm 1.2cm},clip,width=0.45\textwidth]{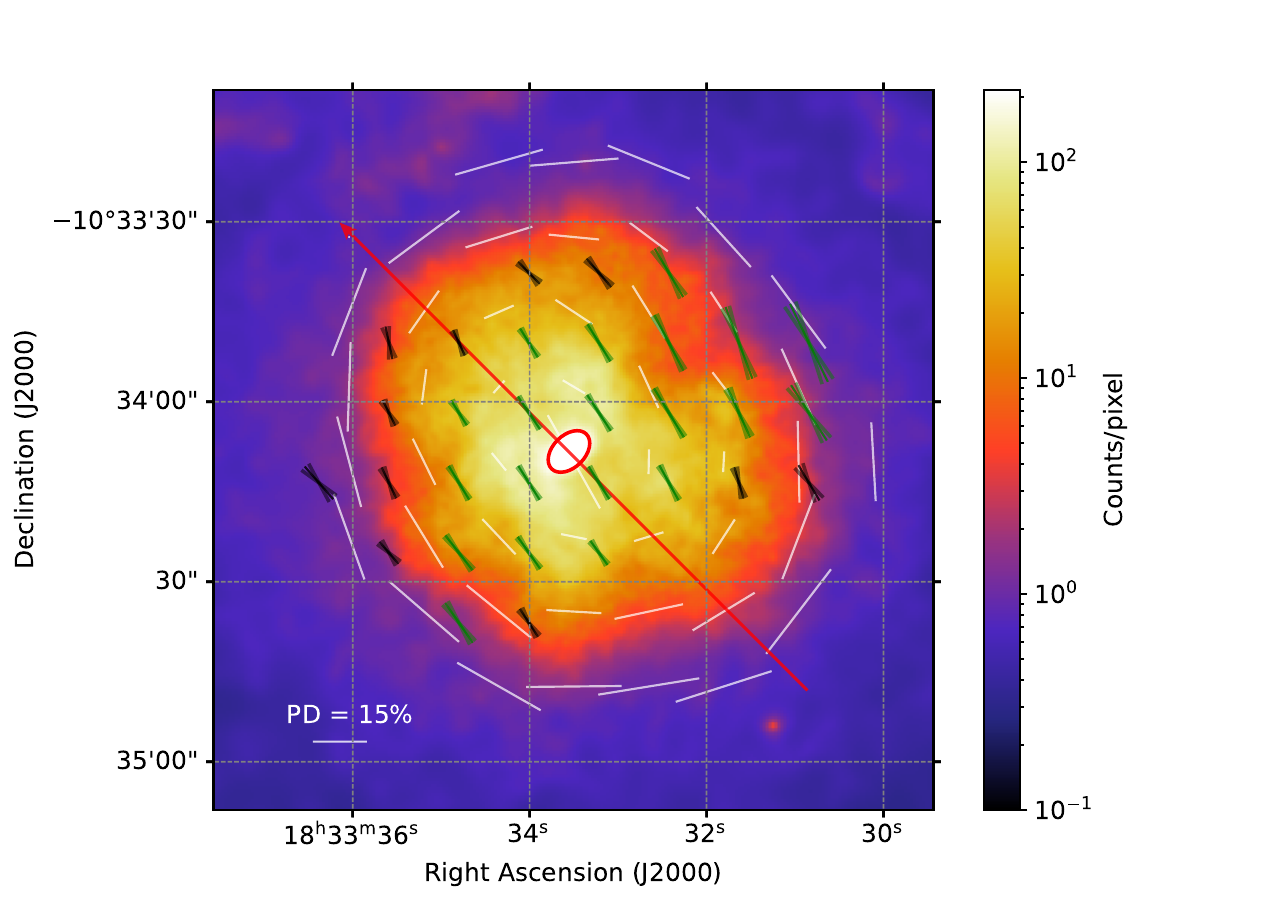}
    \caption{Leakage-subtracted polarization map of the PWN as determined from the space-resolved analysis (2--8~keV) in a 0.1944\arcmin\ grid. Each pixel is correlated with its neighbors as described in the text. Length and orientation of the arrows represent the measured local PD and PA. The black and green lines are the IXPE polarization measurements with a $> 2\sigma$ and $> 3\sigma$ significance, respectively, with additional two sets of lines of the same colors (but different orientation) indicating the associated $1\sigma$ uncertainty on the PA. The white, tangentially-distributed lines are the polarization structure observed in radio~\citep{2022ApJ...930....1L}, corresponding to a radial magnetic field configuration. At the bottom left corner is the reference arrow length for PD = $15\%$. The red arrow represents the projected spin-axis direction of PSR J1833$-$1034 of $\sim 45\degree$~\citep{2008ApJ...673..411N}. The red ellipse roughly corresponds to the region of the inner compact nebula where high level of polarization was measured in IR~\citep{2012A&A...542A..12Z}. Background image is the deep \chandra{} image zoomed in over the PWN.}
    \label{fig:pmap}
\end{figure}

Figure~\ref{fig:pmap} shows the detected IXPE polarization pattern, overlaid on top of the deep \chandra{} count map of \src{} zoomed over the PWN region. The effect of polarization leakage has been estimated and subtracted with \texttt{LeakageLib}, using the same \chandra{} image to model the source.
The measured polarization is visualized using a set of segments,
one per pixel, whose length and inclination represent the local PD (in scale) and PA (measured east of north). The segment color indicates the level of significance of the detection: green corresponds to measurements with significance $> 3\sigma$, black to those between $2\sigma$ and $3\sigma$. For comparison, white segments represent the polarization pattern measured in the radio band~\citep{2022ApJ...930....1L}. Measurements with polarization consistent with zero within $2\sigma$ are not shown. As shown in Fig.~\ref{fig:pmap}, the detected pattern in the central region of the PWN, where the statistical significance is highest, is roughly consistent with a constant PD of 11--12\% at an angle of 30--35\degree.  In the outer regions, the detection is generally less significant due to lower photon statistics. While the measured PD and PA values in these regions appear to vary slightly across the nebula, they are most likely still consistent with a uniform polarization pattern when accounting for their larger uncertainties and the systematics introduced by the leakage correction. This apparent variation, along with the inclusion of regions where no significant polarization was observed---such as the SW portion of the PWN (bottom right)--- helps explain why the space-integrated analysis (Sect.~\ref{sec:spint}), which includes the entire PWN, yielded a lower average PD. The only notable exception is the NW region (top right), where the PD increases progressively, reaching values as high as 20--30\% with a PA around 20--25\degree. No significant polarization was detected outside the PWN. 

\section{Discussion} \label{sec:dis}
Polarization studies provide a powerful diagnostic tool for probing the magnetic field structure and particle acceleration mechanisms in PWNe. In the case of \src{}, IXPE observations allow us to explore the polarization properties of the X-ray-emitting electrons and compare them with those observed in the radio band. This comparison offers crucial insights into the evolution of the PWN’s magnetic field and its interaction with the surrounding SNR environment.

The model that assumes the same outer polarization structure observed in the radio band—where the magnetic field is predominantly radial with respect to the pulsar—predicts that, in an annulus with inner and outer radii of $30$\arcsec\ and $60$\arcsec, respectively, the PA should be tangential, with a background-subtracted and leakage-corrected PD~$\simeq 20\%\pm5\%$. Using the \texttt{xpstokesalign} tool of \ixpeobssim{}, we aligned the reconstructed Stokes parameters of this region to a tangential (radio-like) polarization model by rotating the Stokes vectors event by event, based on their spatial position. Despite an MDP of $\sim10\%$, we did not find evidence of significant polarization.
This result strongly suggests that the magnetic field structures traced by the radio- and X-ray-emitting electrons are different. In this respect, \src{} appears more similar to the Crab Nebula than to Vela, where the radio and X-ray polarization structures are well aligned~\citep{Xie22,Liu23}. Interestingly, both \src{} and the Crab Nebula are believed to be in the \textit{free-expansion} phase, where the PWN expands into the cold SNR ejecta~\citep{Bucciantini_Blondin+03a}, and strong Rayleigh-Taylor instabilities shape the outer regions~\citep{Jun98a,Bucciantini_Amato+04a}. Vela, on the other hand, is likely in a later evolutionary phase, where interaction with the SNR reverse shock is modifying the structure and dynamics of the nebula~\citep{van-der-Swaluw_Achterberg+03a}.

Given that \src{} is at the resolution limit of IXPE, we cannot directly determine which specific nebular structures contribute most to the observed polarization. In particular, two competing scenarios remain indistinguishable: (i) a highly polarized central torus dominating the polarized emission of the source, as suggested by infrared observations~\citep{2012A&A...542A..12Z}, or (ii) a bulk PWN magnetic field that is more uniform but with a lower degree of polarization ($\sim 10\%$). However, the latter scenario is inconsistent with radio observations, which show a clearly radial magnetic field pattern~\citep{2022ApJ...930....1L}.

\begin{deluxetable}{lcr}
\label{tab:model}
\tabletypesize{\scriptsize}
\tablewidth{0pt}
\tablecaption{Torus and nebula models.}
\tablehead{
\colhead{Nebula Model} & \colhead{Torus PD} & \colhead{Torus PA} }
    \startdata 
        Unpolarized & $44\%\pm 7\%$ & $34^\circ \pm 5^\circ$  \\
        Radio-like polarized  & $37\%\pm7\%$ & $29^\circ \pm 5^\circ$ 
    \enddata
    \tablecomments{Different model configurations for the inner torus and nebula that reproduce the polarization properties measured by IXPE in the central region of the PWN. The intrinsic polarization values for the torus are obtained through model fitting using \ixpeobssim\ simulations for each nebula configuration.}
\end{deluxetable}
To evaluate the possible contribution of a highly polarized torus, we simulated two different scenarios using \ixpeobssim: (i) a uniformly polarized torus embedded in an unpolarized nebula, and (ii) the same torus within a nebula that follows the polarization structure and degree observed in the radio band at 5~GHz~\citep{2022ApJ...930....1L}. In both cases, the polarized torus is modeled as an elliptical region ($3.3$\arcsec$\times 7.0$\arcsec) with uniform polarization, centered on the \chandra{} X-ray intensity peak and oriented with a position angle on the plane of the sky of $30^\circ$, consistent with the measured PA and predictions from magneto-hydrodynamic models. We varied the torus PD and PA in the simulations until we achieved consistency with the polarization properties measured in the central region of the PWN (space-integrated \texttt{PCUBE} analysis within a radius $20$\arcsec\ from the PSR: PD~$=12\%\pm 2\%$, PA~$=34^\circ \pm 5^\circ$). This region was chosen because it provides the highest polarization significance while minimizing leakage effects, which are not simulated by \ixpeobssim\ (see Appendix~\ref{sec:app_sim}). The results of these simulations for the two model configurations are shown in Table~\ref{tab:model}. Both models yield large uncertainties in the intrinsic polarization properties of the torus, and neither scenario is strongly favored, though the unpolarized nebula case appears somewhat closer to the infrared observations. In either case, the results are consistent with previous IXPE studies of other PWNe~\citep{Bucciantini23a,Romani23a,Xie22}, suggesting that even in the case of \src{}, close to the termination shock—where particles are likely being accelerated to high energies—turbulence remains moderate ($\delta B / B$ at most $\sim 0.8-1$), indicating a relatively ordered magnetic field.

The PA measured by IXPE for \src{} in the space-integrated analysis differs by approximately $3\sigma$ from the spin-axis orientation of PSR J1833$-$1034 inferred from morphological modeling of the X-ray torus~\citep{2008ApJ...673..411N}. This discrepancy may be explained by the fact that the IXPE measurement reflects the integrated polarization from the entire PWN, including contributions from the more extended outer regions, rather than isolating the compact torus. As demonstrated by our simulations (Table~\ref{tab:model}), even small residual polarization from the outer nebula or deviations from perfect axisymmetry can bias the inferred PA away from the spin-axis direction. A similar effect has been observed in the Crab nebula~\citep{Bucciantini23a}, where non-axisymmetric surface brightness in the X-ray torus produces a measurable offset between PA and the PWN symmetry axis. Given these considerations and the modest statistical tension, we regard the measured PA and spin-axis orientation as broadly consistent within uncertainties.

Our findings suggest that a two-zone model for \src{} is probably preferred: a central bright X-ray torus dominated by a globally toroidal magnetic field, surrounded by an outer layer affected by Rayleigh-Taylor instabilities, similar to the filamentary network seen in the Crab Nebula, which stretches the magnetic field radially. However, the overall low integrated polarization of \src{} distinguishes this source from other well-studied PWNe like the Crab, Vela, and MSH 15$-$52. This fact, along with differences in the radio polarization pattern and the absence of a strong jet-torus structure, suggests that X-ray-emitting particles may populate a larger region of the PWN where turbulence is stronger or where magnetic field geometry varies significantly due to internal dynamics or enhanced diffusion.

\begin{acknowledgments}
The Imaging X-ray Polarimetry Explorer (IXPE) is a joint US and Italian mission.  The US contribution is supported by the National Aeronautics and Space Administration (NASA) and led and managed by its Marshall Space Flight Center (MSFC), with industry partner Ball Aerospace (contract NNM15AA18C).  The Italian contribution is supported by the Italian Space Agency (Agenzia Spaziale Italiana, ASI) through contract ASI-OHBI-2022-13-I.0, agreements ASI-INAF-2022-19-HH.0 and ASI-INFN-2017.13-H0, and its Space Science Data Center (SSDC) with agreements ASI-INAF-2022-14-HH.0 and ASI-INFN 2021-43-HH.0, and by the Istituto Nazionale di Astrofisica (INAF) and the Istituto Nazionale di Fisica Nucleare (INFN) in Italy.  This research used data products provided by the IXPE Team (MSFC, SSDC, INAF, and INFN) and distributed with additional software tools by the High-Energy Astrophysics Science Archive Research Center (HEASARC), at NASA Goddard Space Flight Center (GSFC). R.F., E.Co., A.D.M., S.F., F.L.M., F.Mu., and P.So. are partially supported by MAECI with grant CN24GR08 ``GRBAXP: Guangxi Rome Bilateral Agreement for X-ray Polarimetry in Astrophysics''. C.-Y.N. is supported by a GRF grant of the Hong Kong Government under HKU 17304524.  N.B. was supported by the INAF MiniGrant ``PWNnumpol - Numerical Studies of Pulsar Wind Nebulae in The Light of IXPE.'' F.X. is supported by National Natural Science Foundation of China (grant Nos. 12373041 and 12422306), and Bagui Scholars Program (XF). I.L. was funded by the European Union ERC-2022-STG - BOOTES - 101076343. Views and opinions expressed are however those of the author(s) only and do not necessarily reflect those of the European Union or the European Research Council Executive Agency. Neither the European Union nor the granting authority can be held responsible for them.
\end{acknowledgments}





%
\facilities{IXPE, \chandra{}}

\software{\ixpeobssim~\citep{Baldini2022}, \ixpesim~\citep{dilalla2019}, \texttt{3ML}~\citep{2015arXiv150708343V}, \texttt{LeakageLib}~\citep{Dinsmore2024}}

\bibliography{g21_rev}{}

\begin{thebibliography}{}
\expandafter\ifx\csname natexlab\endcsname\relax\def\natexlab#1{#1}\fi
\providecommand{\url}[1]{\href{#1}{#1}}
\providecommand{\dodoi}[1]{doi:~\href{http://doi.org/#1}{\nolinkurl{#1}}}
\providecommand{\doeprint}[1]{\href{http://ascl.net/#1}{\nolinkurl{http://ascl.net/#1}}}
\providecommand{\doarXiv}[1]{\href{https://arxiv.org/abs/#1}{\nolinkurl{https://arxiv.org/abs/#1}}}

\bibitem[{{Abdo} {et~al.}(2013){Abdo}, {Ajello}, {Allafort}, {Baldini}, {Ballet}, {Barbiellini}, {Baring}, {Bastieri}, {Belfiore}, {Bellazzini}, {Bhattacharyya}, {Bissaldi}, {Bloom}, {Bonamente}, {Bottacini}, {Brandt}, {Bregeon}, {Brigida}, {Bruel}, {Buehler}, {Burgay}, {Burnett}, {Busetto}, {Buson}, {Caliandro}, {Cameron}, {Camilo}, {Caraveo}, {Casandjian}, {Cecchi}, {{\c{C}}elik}, {Charles}, {Chaty}, {Chaves}, {Chekhtman}, {Chen}, {Chiang}, {Chiaro}, {Ciprini}, {Claus}, {Cognard}, {Cohen-Tanugi}, {Cominsky}, {Conrad}, {Cutini}, {D'Ammando}, {de Angelis}, {DeCesar}, {De Luca}, {den Hartog}, {de Palma}, {Dermer}, {Desvignes}, {Digel}, {Di Venere}, {Drell}, {Drlica-Wagner}, {Dubois}, {Dumora}, {Espinoza}, {Falletti}, {Favuzzi}, {Ferrara}, {Focke}, {Franckowiak}, {Freire}, {Funk}, {Fusco}, {Gargano}, {Gasparrini}, {Germani}, {Giglietto}, {Giommi}, {Giordano}, {Giroletti}, {Glanzman}, {Godfrey}, {Gotthelf}, {Grenier}, {Grondin}, {Grove}, {Guillemot}, {Guiriec}, {Hadasch}, {Hanabata}, {Harding}, {Hayashida},
  {Hays}, {Hessels}, {Hewitt}, {Hill}, {Horan}, {Hou}, {Hughes}, {Jackson}, {Janssen}, {Jogler}, {J{\'o}hannesson}, {Johnson}, {Johnson}, {Johnson}, {Johnson}, {Johnston}, {Kamae}, {Kataoka}, {Keith}, {Kerr}, {Kn{\"o}dlseder}, {Kramer}, {Kuss}, {Lande}, {Larsson}, {Latronico}, {Lemoine-Goumard}, {Longo}, {Loparco}, {Lovellette}, {Lubrano}, {Lyne}, {Manchester}, {Marelli}, {Massaro}, {Mayer}, {Mazziotta}, {McEnery}, {McLaughlin}, {Mehault}, {Michelson}, {Mignani}, {Mitthumsiri}, {Mizuno}, {Moiseev}, {Monzani}, {Morselli}, {Moskalenko}, {Murgia}, {Nakamori}, {Nemmen}, {Nuss}, {Ohno}, {Ohsugi}, {Orienti}, {Orlando}, {Ormes}, {Paneque}, {Panetta}, {Parent}, {Perkins}, {Pesce-Rollins}, {Pierbattista}, {Piron}, {Pivato}, {Pletsch}, {Porter}, {Possenti}, {Rain{\`o}}, {Rando}, {Ransom}, {Ray}, {Razzano}, {Rea}, {Reimer}, {Reimer}, {Renault}, {Reposeur}, {Ritz}, {Romani}, {Roth}, {Rousseau}, {Roy}, {Ruan}, {Sartori}, {Saz Parkinson}, {Scargle}, {Schulz}, {Sgr{\`o}}, {Shannon}, {Siskind}, {Smith}, {Spandre},
  {Spinelli}, {Stappers}, {Strong}, {Suson}, {Takahashi}, {Thayer}, {Thayer}, {Theureau}, {Thompson}, {Thorsett}, {Tibaldo}, {Tibolla}, {Tinivella}, {Torres}, {Tosti}, {Troja}, {Uchiyama}, {Usher}, {Vandenbroucke}, {Vasileiou}, {Venter}, {Vianello}, {Vitale}, {Wang}, {Weltevrede}, {Winer}, {Wolff}, {Wood}, {Wood}, {Wood}, \& {Yang}}]{2013ApJS..208...17A}
{Abdo}, A.~A., {Ajello}, M., {Allafort}, A., {et~al.} 2013, \apjs, 208, 17, \dodoi{10.1088/0067-0049/208/2/17}

\bibitem[{Agostinelli {et~al.}(2003)}]{GEANT4:2002zbu}
Agostinelli, S., {et~al.} 2003, Nucl. Instrum. Meth. A, 506, 250, \dodoi{10.1016/S0168-9002(03)01368-8}

\bibitem[{{Baldini} {et~al.}(2021){Baldini}, {Barbanera}, {Bellazzini}, {Bonino}, {Borotto}, {Brez}, {Caporale}, {Cardelli}, {Castellano}, {Ceccanti}, {Citraro}, {Di Lalla}, {Latronico}, {Lucchesi}, {Magazz{\`u}}, {Magazz{\`u}}, {Maldera}, {Manfreda}, {Marengo}, {Marrocchesi}, {Mereu}, {Minuti}, {Mosti}, {Nasimi}, {Nuti}, {Oppedisano}, {Orsini}, {Pesce-Rollins}, {Pinchera}, {Profeti}, {Sgr{\`o}}, {Spandre}, {Tardiola}, {Zanetti}, {Amici}, {Andersson}, {Attin{\`a}}, {Bachetti}, {Baumgartner}, {Brienza}, {Carpentiero}, {Castronuovo}, {Cavalli}, {Cavazzuti}, {Centrone}, {Costa}, {D'Alba}, {D'Amico}, {Del Monte}, {Di Cosimo}, {Di Marco}, {Di Persio}, {Donnarumma}, {Evangelista}, {Fabiani}, {Ferrazzoli}, {Kitaguchi}, {La Monaca}, {Lefevre}, {Loffredo}, {Lorenzi}, {Mangraviti}, {Matt}, {Meilahti}, {Morbidini}, {Muleri}, {Nakano}, {Negri}, {Nenonen}, {O'Dell}, {Perri}, {Piazzolla}, {Pieraccini}, {Pilia}, {Puccetti}, {Ramsey}, {Rankin}, {Ratheesh}, {Rubini}, {Santoli}, {Sarra}, {Scalise}, {Sciortino}, {Soffitta},
  {Tamagawa}, {Tennant}, {Tobia}, {Trois}, {Uchiyama}, {Vimercati}, {Weisskopf}, {Xie}, {Zanetti}, \& {Zhou}}]{2021APh...13302628B}
{Baldini}, L., {Barbanera}, M., {Bellazzini}, R., {et~al.} 2021, Astroparticle Physics, 133, 102628, \dodoi{10.1016/j.astropartphys.2021.102628}

\bibitem[{{Baldini} {et~al.}(2022){Baldini}, {Bucciantini}, {Di Lalla}, {Ehlert}, {Manfreda}, {Negro}, {Omodei}, {Pesce-Rollins}, {Sgr{\`o}}, \& {Silvestri}}]{Baldini2022}
{Baldini}, L., {Bucciantini}, N., {Di Lalla}, N., {et~al.} 2022, SoftwareX, 19, 101194, \dodoi{10.1016/j.softx.2022.101194}

\bibitem[{{Becker} \& {Kundu}(1976)}]{1976ApJ...204..427B}
{Becker}, R.~H., \& {Kundu}, M.~R. 1976, \apj, 204, 427, \dodoi{10.1086/154186}

\bibitem[{{Becker} \& {Szymkowiak}(1981)}]{1981ApJ...248L..23B}
{Becker}, R.~H., \& {Szymkowiak}, A.~E. 1981, \apjl, 248, L23, \dodoi{10.1086/183615}

\bibitem[{{Bellazzini} {et~al.}(2006){Bellazzini}, {Angelini}, {Baldini}, {Bitti}, {Brez}, {Cavalca}, {Del Prete}, {Kuss}, {Latronico}, {Omodei}, {Pinchera}, {Massai}, {Minuti}, {Razzano}, {Sgro}, {Spandre}, {Tenze}, {Costa}, \& {Soffitta}}]{2006NIMPA.560..425B}
{Bellazzini}, R., {Angelini}, F., {Baldini}, L., {et~al.} 2006, Nuclear Instruments and Methods in Physics Research A, 560, 425, \dodoi{10.1016/j.nima.2006.01.046}

\bibitem[{{Bietenholz} \& {Bartel}(2008)}]{2008MNRAS.386.1411B}
{Bietenholz}, M.~F., \& {Bartel}, N. 2008, \mnras, 386, 1411, \dodoi{10.1111/j.1365-2966.2008.13058.x}

\bibitem[{{Bietenholz} {et~al.}(2011){Bietenholz}, {Matheson}, {Safi-Harb}, {Brogan}, \& {Bartel}}]{2011MNRAS.412.1221B}
{Bietenholz}, M.~F., {Matheson}, H., {Safi-Harb}, S., {Brogan}, C., \& {Bartel}, N. 2011, \mnras, 412, 1221, \dodoi{10.1111/j.1365-2966.2010.17981.x}

\bibitem[{{Bocchino} {et~al.}(2005){Bocchino}, {van der Swaluw}, {Chevalier}, \& {Bandiera}}]{2005A&A...442..539B}
{Bocchino}, F., {van der Swaluw}, E., {Chevalier}, R., \& {Bandiera}, R. 2005, \aap, 442, 539, \dodoi{10.1051/0004-6361:20052870}

\bibitem[{{Bucciantini} {et~al.}(2004){Bucciantini}, {Amato}, {Bandiera}, {Blondin}, \& {Del Zanna}}]{Bucciantini_Amato+04a}
{Bucciantini}, N., {Amato}, E., {Bandiera}, R., {Blondin}, J.~M., \& {Del Zanna}, L. 2004, \aap, 423, 253, \dodoi{10.1051/0004-6361:20040360}

\bibitem[{{Bucciantini} {et~al.}(2003){Bucciantini}, {Blondin}, {Del Zanna}, \& {Amato}}]{Bucciantini_Blondin+03a}
{Bucciantini}, N., {Blondin}, J.~M., {Del Zanna}, L., \& {Amato}, E. 2003, \aap, 405, 617, \dodoi{10.1051/0004-6361:20030624}

\bibitem[{{Bucciantini} {et~al.}(2023{\natexlab{a}}){Bucciantini}, {Di Lalla}, {Romani}, {Silvestri}, {Negro}, {Baldini}, {Tennant}, \& {Manfreda}}]{Bucciantini2023Leakage}
{Bucciantini}, N., {Di Lalla}, N., {Romani}, R.~W.~R., {et~al.} 2023{\natexlab{a}}, \aap, 672, A66, \dodoi{10.1051/0004-6361/202245744}

\bibitem[{{Bucciantini} {et~al.}(2023{\natexlab{b}}){Bucciantini}, {Ferrazzoli}, {Bachetti}, {Rankin}, {Di Lalla}, {Sgr{\`o}}, {Omodei}, {Kitaguchi}, {Mizuno}, {Gunji}, {Watanabe}, {Baldini}, {Slane}, {Weisskopf}, {Romani}, {Possenti}, {Marshall}, {Silvestri}, {Pacciani}, {Negro}, {Muleri}, {de O{\~n}a Wilhelmi}, {Xie}, {Heyl}, {Pesce-Rollins}, {Wong}, {Pilia}, {Agudo}, {Antonelli}, {Baumgartner}, {Bellazzini}, {Bianchi}, {Bongiorno}, {Bonino}, {Brez}, {Capitanio}, {Castellano}, {Cavazzuti}, {Chen}, {Ciprini}, {Costa}, {De Rosa}, {Del Monte}, {Di Gesu}, {Di Marco}, {Donnarumma}, {Doroshenko}, {Dov{\v{c}}iak}, {Ehlert}, {Enoto}, {Evangelista}, {Fabiani}, {Garcia}, {Hayashida}, {Iwakiri}, {Jorstad}, {Kaaret}, {Karas}, {Kislat}, {Kolodziejczak}, {Krawczynski}, {La Monaca}, {Latronico}, {Liodakis}, {Maldera}, {Manfreda}, {Marin}, {Marinucci}, {Marscher}, {Massaro}, {Matt}, {Mitsuishi}, {Ng}, {O'Dell}, {Oppedisano}, {Papitto}, {Pavlov}, {Peirson}, {Perri}, {Petrucci}, {Poutanen}, {Puccetti}, {Ramsey}, {Ratheesh},
  {Roberts}, {Soffitta}, {Spandre}, {Swartz}, {Tamagawa}, {Tavecchio}, {Taverna}, {Tawara}, {Tennant}, {Thomas}, {Tombesi}, {Trois}, {Tsygankov}, {Turolla}, {Vink}, {Wu}, \& {Zane}}]{Bucciantini23a}
{Bucciantini}, N., {Ferrazzoli}, R., {Bachetti}, M., {et~al.} 2023{\natexlab{b}}, Nature Astronomy, 7, 602, \dodoi{10.1038/s41550-023-01936-8}

\bibitem[{{Camilo} {et~al.}(2006){Camilo}, {Ransom}, {Gaensler}, {Slane}, {Lorimer}, {Reynolds}, {Manchester}, \& {Murray}}]{2006ApJ...637..456C}
{Camilo}, F., {Ransom}, S.~M., {Gaensler}, B.~M., {et~al.} 2006, \apj, 637, 456, \dodoi{10.1086/498386}

\bibitem[{{Cibrario} {et~al.}(2023){Cibrario}, {Negro}, {Moriakov}, {Bonino}, {Baldini}, {Di Lalla}, {Latronico}, {Maldera}, {Manfreda}, {Omodei}, {Sgr{\'o}}, \& {Tugliani}}]{2023A&A...674A.107C}
{Cibrario}, N., {Negro}, M., {Moriakov}, N., {et~al.} 2023, \aap, 674, A107, \dodoi{10.1051/0004-6361/202346302}

\bibitem[{Cibrario {et~al.}(2025)Cibrario, Negro, Bonino, Moriakov, Baldini, Di~Lalla, Di~Marco, Fabiani, Frassá, Gorgi, La~Monaca, Latronico, Maldera, Manfreda, Muleri, Omodei, Rankin, Sgró, Silvestri, Soffitta, \& Tugliani}]{hybrid_2}
Cibrario, N., Negro, M., Bonino, R., {et~al.} 2025, The Astrophysical Journal, 984, 171, \dodoi{10.3847/1538-4357/adc92c}

\bibitem[{{Di Lalla}(2019)}]{dilalla2019}
{Di Lalla}, N. 2019, PhD thesis, University of Pisa.
\newblock \url{https://etd.adm.unipi.it/t/etd-04042019-100412}

\bibitem[{{Di Marco} {et~al.}(2023){Di Marco}, {Soffitta}, {Costa}, {Ferrazzoli}, {La Monaca}, {Rankin}, {Ratheesh}, {Xie}, {Baldini}, {Del Monte}, {Ehlert}, {Fabiani}, {Kim}, {Muleri}, {O'Dell}, {Ramsey}, {Rubini}, {Sgr{\`o}}, {Silvestri}, {Tennant}, \& {Weisskopf}}]{2023AJ....165..143D}
{Di Marco}, A., {Soffitta}, P., {Costa}, E., {et~al.} 2023, \aj, 165, 143, \dodoi{10.3847/1538-3881/acba0f}

\bibitem[{{Dinsmore} \& {Romani}(2024)}]{Dinsmore2024}
{Dinsmore}, J.~T., \& {Romani}, R.~W. 2024, \apj, 962, 183, \dodoi{10.3847/1538-4357/ad2065}

\bibitem[{{Furst} {et~al.}(1988){Furst}, {Handa}, {Morita}, {Reich}, {Reich}, \& {Sofue}}]{1988PASJ...40..347F}
{Furst}, E., {Handa}, T., {Morita}, K., {et~al.} 1988, \pasj, 40, 347

\bibitem[{{Gaensler} \& {Slane}(2006)}]{Gaensler2006}
{Gaensler}, B.~M., \& {Slane}, P.~O. 2006, \araa, 44, 17, \dodoi{10.1146/annurev.astro.44.051905.092528}

\bibitem[{{Guest} {et~al.}(2019){Guest}, {Safi-Harb}, \& {Tang}}]{2019MNRAS.482.1031G}
{Guest}, B.~T., {Safi-Harb}, S., \& {Tang}, X. 2019, \mnras, 482, 1031, \dodoi{10.1093/mnras/sty2635}

\bibitem[{{Gupta} {et~al.}(2005){Gupta}, {Mitra}, {Green}, \& {Acharyya}}]{2005CSci...89..853G}
{Gupta}, Y., {Mitra}, D., {Green}, D.~A., \& {Acharyya}, A. 2005, Current Science, 89, 853, \dodoi{10.48550/arXiv.astro-ph/0508257}

\bibitem[{{Hewish} {et~al.}(1968){Hewish}, {Bell}, {Pilkington}, {Scott}, \& {Collins}}]{1968Natur.217..709H}
{Hewish}, A., {Bell}, S.~J., {Pilkington}, J.~D.~H., {Scott}, P.~F., \& {Collins}, R.~A. 1968, \nat, 217, 709, \dodoi{10.1038/217709a0}

\bibitem[{{Jun}(1998)}]{Jun98a}
{Jun}, B.-I. 1998, \apj, 499, 282, \dodoi{10.1086/305627}

\bibitem[{{Kaastra} \& {Bleeker}(2016)}]{2016A&A...587A.151K}
{Kaastra}, J.~S., \& {Bleeker}, J.~A.~M. 2016, \aap, 587, A151, \dodoi{10.1051/0004-6361/201527395}

\bibitem[{Kislat {et~al.}(2015)Kislat, Clark, Beilicke, \& Krawczynski}]{Stokes}
Kislat, F., Clark, B., Beilicke, M., \& Krawczynski, H. 2015, Astroparticle Physics, 68, 45, \dodoi{https://doi.org/10.1016/j.astropartphys.2015.02.007}

\bibitem[{{Lai} {et~al.}(2022){Lai}, {Ng}, \& {Bucciantini}}]{2022ApJ...930....1L}
{Lai}, P. C.~W., {Ng}, C.~Y., \& {Bucciantini}, N. 2022, \apj, 930, 1, \dodoi{10.3847/1538-4357/ac63b1}

\bibitem[{{Liu} {et~al.}(2023){Liu}, {Xie}, {Liu}, {Ng}, {Bucciantini}, {Romani}, {Weisskopf}, {Costa}, {Di Marco}, {La Monaca}, {Muleri}, {Soffitta}, {Deng}, {Meng}, \& {Liang}}]{Liu23}
{Liu}, K., {Xie}, F., {Liu}, Y.-h., {et~al.} 2023, \apjl, 959, L2, \dodoi{10.3847/2041-8213/ad0bfc}

\bibitem[{Matheson \& Safi-Harb(2005)}]{MATHESON20051099}
Matheson, H., \& Safi-Harb, S. 2005, Advances in Space Research, 35, 1099, \dodoi{https://doi.org/10.1016/j.asr.2005.04.050}

\bibitem[{{Matheson} \& {Safi-Harb}(2010)}]{2010ApJ...724..572M}
{Matheson}, H., \& {Safi-Harb}, S. 2010, \apj, 724, 572, \dodoi{10.1088/0004-637X/724/1/572}

\bibitem[{{Negro} {et~al.}(2023){Negro}, {Di Lalla}, {Omodei}, {Veres}, {Silvestri}, {Manfreda}, {Burns}, {Baldini}, {Costa}, {Ehlert}, {Kennea}, {Liodakis}, {Marshall}, {Mereghetti}, {Middei}, {Muleri}, {O'Dell}, {Roberts}, {Romani}, {Sgr{\'o}}, {Terashima}, {Tiengo}, {Viscolo}, {Di Marco}, {La Monaca}, {Latronico}, {Matt}, {Perri}, {Puccetti}, {Poutanen}, {Ratheesh}, {Rogantini}, {Slane}, {Soffitta}, {Lindfors}, {Nilsson}, {Kasikov}, {Marscher}, {Tavecchio}, {Cibrario}, {Gunji}, {Malacaria}, {Paggi}, {Yang}, {Zane}, {Weisskopf}, {Agudo}, {Antonelli}, {Bachetti}, {Baumgartner}, {Bellazzini}, {Bianchi}, {Bongiorno}, {Bonino}, {Brez}, {Bucciantini}, {Capitanio}, {Castellano}, {Cavazzuti}, {Chen}, {Ciprini}, {De Rosa}, {Del Monte}, {Di Gesu}, {Donnarumma}, {Doroshenko}, {Dovc̆iak}, {Enoto}, {Evangelista}, {Fabiani}, {Ferrazzoli}, {Garcia}, {Hayashida}, {Heyl}, {Iwakiri}, {Jorstad}, {Kaaret}, {Karas}, {Kislat}, {Kitaguchi}, {Kolodziejczak}, {Krawczynski}, {Maldera}, {Marin}, {Marinucci}, {Mitsuishi}, {Mizuno},
  {Ng}, {Oppedisano}, {Papitto}, {Pavlov}, {Peirson}, {Pesce-Rollins}, {Petrucci}, {Pilia}, {Possenti}, {Ramsey}, {Rankin}, {Spandre}, {Swartz}, {Tamagawa}, {Taverna}, {Tawara}, {Tennant}, {Thomas}, {Tombesi}, {Trois}, {Tsygankov}, {Turolla}, {Vink}, {Wu}, \& {Xie}}]{2023ApJ...946L..21N}
{Negro}, M., {Di Lalla}, N., {Omodei}, N., {et~al.} 2023, \apjl, 946, L21, \dodoi{10.3847/2041-8213/acba17}

\bibitem[{{Ng} \& {Romani}(2008)}]{2008ApJ...673..411N}
{Ng}, C.~Y., \& {Romani}, R.~W. 2008, \apj, 673, 411, \dodoi{10.1086/523935}

\bibitem[{{Romani} {et~al.}(2023){Romani}, {Wong}, {Di Lalla}, {Omodei}, {Xie}, {Ng}, {Ferrazzoli}, {Di Marco}, {Bucciantini}, {Pilia}, {Slane}, {Weisskopf}, {Johnston}, {Burgay}, {Wei}, {Yang}, {Zhang}, {Antonelli}, {Bachetti}, {Baldini}, {Baumgartner}, {Bellazzini}, {Bianchi}, {Bongiorno}, {Bonino}, {Brez}, {Capitanio}, {Castellano}, {Cavazzuti}, {Chen}, {Cibrario}, {Ciprini}, {Costa}, {De Rosa}, {Del Monte}, {Di Gesu}, {Donnarumma}, {Doroshenko}, {Dov{\v{c}}iak}, {Ehlert}, {Enoto}, {Evangelista}, {Fabiani}, {Garcia}, {Gunji}, {Hayashida}, {Heyl}, {Iwakiri}, {Liodakis}, {Kaaret}, {Karas}, {Kim}, {Kitaguchi}, {Kolodziejczak}, {Krawczynski}, {La Monaca}, {Latronico}, {Madejski}, {Maldera}, {Manfreda}, {Marin}, {Marinucci}, {Marscher}, {Marshall}, {Massaro}, {Matt}, {Middei}, {Mitsuishi}, {Mizuno}, {Muleri}, {Negro}, {O'Dell}, {Oppedisano}, {Pacciani}, {Papitto}, {Pavlov}, {Perri}, {Pesce-Rollins}, {Petrucci}, {Possenti}, {Poutanen}, {Puccetti}, {Ramsey}, {Rankin}, {Ratheesh}, {Roberts}, {Sgr{\'o}},
  {Soffitta}, {Spandre}, {Swartz}, {Tamagawa}, {Tavecchio}, {Taverna}, {Tawara}, {Tennant}, {Thomas}, {Tombesi}, {Trois}, {Tsygankov}, {Turolla}, {Vink}, {Wu}, \& {Zane}}]{Romani23a}
{Romani}, R.~W., {Wong}, J., {Di Lalla}, N., {et~al.} 2023, \apj, 957, 23, \dodoi{10.3847/1538-4357/acfa02}

\bibitem[{{Safi-Harb} {et~al.}(2001){Safi-Harb}, {Harrus}, {Petre}, {Pavlov}, {Koptsevich}, \& {Sanwal}}]{2001ApJ...561..308S}
{Safi-Harb}, S., {Harrus}, I.~M., {Petre}, R., {et~al.} 2001, \apj, 561, 308, \dodoi{10.1086/322978}

\bibitem[{{Slane} {et~al.}(2000){Slane}, {Chen}, {Schulz}, {Seward}, {Hughes}, \& {Gaensler}}]{2000ApJ...533L..29S}
{Slane}, P., {Chen}, Y., {Schulz}, N.~S., {et~al.} 2000, \apjl, 533, L29, \dodoi{10.1086/312589}

\bibitem[{{Smith} {et~al.}(2023){Smith}, {Abdollahi}, {Ajello}, {Bailes}, {Baldini}, {Ballet}, {Baring}, {Bassa}, {Gonzalez}, {Bellazzini}, {Berretta}, {Bhattacharyya}, {Bissaldi}, {Bonino}, {Bottacini}, {Bregeon}, {Bruel}, {Burgay}, {Burnett}, {Cameron}, {Camilo}, {Caputo}, {Caraveo}, {Cavazzuti}, {Chiaro}, {Ciprini}, {Clark}, {Cognard}, {Corongiu}, {Orestano}, {Crnogorcevic}, {Cuoco}, {Cutini}, {D'Ammando}, {de Angelis}, {DeCesar}, {De Gaetano}, {de Menezes}, {Deneva}, {de Palma}, {Di Lalla}, {Dirirsa}, {Di Venere}, {Dom{\'\i}nguez}, {Dumora}, {Fegan}, {Ferrara}, {Fiori}, {Fleischhack}, {Flynn}, {Franckowiak}, {Freire}, {Fukazawa}, {Fusco}, {Galanti}, {Gammaldi}, {Gargano}, {Gasparrini}, {Giacchino}, {Giglietto}, {Giordano}, {Giroletti}, {Green}, {Grenier}, {Guillemot}, {Guiriec}, {Gustafsson}, {Harding}, {Hays}, {Hewitt}, {Horan}, {Hou}, {Jankowski}, {Johnson}, {Johnson}, {Johnston}, {Kataoka}, {Keith}, {Kerr}, {Kramer}, {Kuss}, {Latronico}, {Lee}, {Li}, {Li}, {Limyansky}, {Longo}, {Loparco}, {Lorusso},
  {Lovellette}, {Lower}, {Lubrano}, {Lyne}, {Maan}, {Maldera}, {Manchester}, {Manfreda}, {Marelli}, {Mart{\'\i}-Devesa}, {Mazziotta}, {McEnery}, {Mereu}, {Michelson}, {Mickaliger}, {Mitthumsiri}, {Mizuno}, {Moiseev}, {Monzani}, {Morselli}, {Negro}, {Nemmen}, {Nieder}, {Nuss}, {Omodei}, {Orienti}, {Orlando}, {Ormes}, {Palatiello}, {Paneque}, {Panzarini}, {Parthasarathy}, {Persic}, {Pesce-Rollins}, {Pillera}, {Poon}, {Porter}, {Possenti}, {Principe}, {Rain{\`o}}, {Rando}, {Ransom}, {Ray}, {Razzano}, {Razzaque}, {Reimer}, {Reimer}, {Renault-Tinacci}, {Romani}, {S{\'a}nchez-Conde}, {Parkinson}, {Scotton}, {Serini}, {Sgr{\`o}}, {Shannon}, {Sharma}, {Shen}, {Siskind}, {Spandre}, {Spinelli}, {Stappers}, {Stephens}, {Suson}, {Tabassum}, {Tajima}, {Tak}, {Theureau}, {Thompson}, {Tibolla}, {Torres}, {Valverde}, {Venter}, {Wadiasingh}, {Wang}, {Wang}, {Wang}, {Weltevrede}, {Wood}, {Yan}, {Zaharijas}, {Zhang}, \& {Zhu}}]{2023ApJ...958..191S}
{Smith}, D.~A., {Abdollahi}, S., {Ajello}, M., {et~al.} 2023, \apj, 958, 191, \dodoi{10.3847/1538-4357/acee67}

\bibitem[{{Tian} \& {Leahy}(2008)}]{2008MNRAS.391L..54T}
{Tian}, W.~W., \& {Leahy}, D.~A. 2008, \mnras, 391, L54, \dodoi{10.1111/j.1745-3933.2008.00557.x}

\bibitem[{{Tsujimoto} {et~al.}(2011){Tsujimoto}, {Guainazzi}, {Plucinsky}, {Beardmore}, {Ishida}, {Natalucci}, {Posson-Brown}, {Read}, {Saxton}, \& {Shaposhnikov}}]{2011A&A...525A..25T}
{Tsujimoto}, M., {Guainazzi}, M., {Plucinsky}, P.~P., {et~al.} 2011, \aap, 525, A25, \dodoi{10.1051/0004-6361/201015597}

\bibitem[{{van der Swaluw} {et~al.}(2003){van der Swaluw}, {Achterberg}, {Gallant}, {Downes}, \& {Keppens}}]{van-der-Swaluw_Achterberg+03a}
{van der Swaluw}, E., {Achterberg}, A., {Gallant}, Y.~A., {Downes}, T.~P., \& {Keppens}, R. 2003, \aap, 397, 913, \dodoi{10.1051/0004-6361:20021488}

\bibitem[{{Vianello} {et~al.}(2015){Vianello}, {Lauer}, {Younk}, {Tibaldo}, {Burgess}, {Ayala}, {Harding}, {Hui}, {Omodei}, \& {Zhou}}]{2015arXiv150708343V}
{Vianello}, G., {Lauer}, R.~J., {Younk}, P., {et~al.} 2015, arXiv e-prints, arXiv:1507.08343, \dodoi{10.48550/arXiv.1507.08343}

\bibitem[{Weisskopf {et~al.}(2010)Weisskopf, Elsner, \& O’Dell}]{Weisskopf_2010}
Weisskopf, M.~C., Elsner, R.~F., \& O’Dell, S.~L. 2010, in Space Telescopes and Instrumentation 2010: Ultraviolet to Gamma Ray, ed. M.~Arnaud, S.~S. Murray, \& T.~Takahashi (SPIE), \dodoi{10.1117/12.857357}

\bibitem[{{Weisskopf} {et~al.}(2022){Weisskopf}, {Soffitta}, {Baldini}, {Ramsey}, {O'Dell}, {Romani}, {Matt}, {Deininger}, {Baumgartner}, {Bellazzini}, {Costa}, {Kolodziejczak}, {Latronico}, {Marshall}, {Muleri}, {Bongiorno}, {Tennant}, {Bucciantini}, {Dovciak}, {Marin}, {Marscher}, {Poutanen}, {Slane}, {Turolla}, {Kalinowski}, {Di Marco}, {Fabiani}, {Minuti}, {La Monaca}, {Pinchera}, {Rankin}, {Sgro'}, {Trois}, {Xie}, {Alexander}, {Allen}, {Amici}, {Andersen}, {Antonelli}, {Antoniak}, {Attin{\`a}}, {Barbanera}, {Bachetti}, {Baggett}, {Bladt}, {Brez}, {Bonino}, {Boree}, {Borotto}, {Breeding}, {Brienza}, {Bygott}, {Caporale}, {Cardelli}, {Carpentiero}, {Castellano}, {Castronuovo}, {Cavalli}, {Cavazzuti}, {Ceccanti}, {Centrone}, {Citraro}, {D'Amico}, {D'Alba}, {Di Gesu}, {Del Monte}, {Dietz}, {Di Lalla}, {Persio}, {Dolan}, {Donnarumma}, {Evangelista}, {Ferrant}, {Ferrazzoli}, {Ferrie}, {Footdale}, {Forsyth}, {Foster}, {Garelick}, {Gunji}, {Gurnee}, {Head}, {Hibbard}, {Johnson}, {Kelly}, {Kilaru}, {Lefevre},
  {Roy}, {Loffredo}, {Lorenzi}, {Lucchesi}, {Maddox}, {Magazzu}, {Maldera}, {Manfreda}, {Mangraviti}, {Marengo}, {Marrocchesi}, {Massaro}, {Mauger}, {McCracken}, {McEachen}, {Mize}, {Mereu}, {Mitchell}, {Mitsuishi}, {Morbidini}, {Mosti}, {Nasimi}, {Negri}, {Negro}, {Nguyen}, {Nitschke}, {Nuti}, {Onizuka}, {Oppedisano}, {Orsini}, {Osborne}, {Pacheco}, {Paggi}, {Painter}, {Pavelitz}, {Pentz}, {Piazzolla}, {Perri}, {Pesce-Rollins}, {Peterson}, {Pilia}, {Profeti}, {Puccetti}, {Ranganathan}, {Ratheesh}, {Reedy}, {Root}, {Rubini}, {Ruswick}, {Sanchez}, {Sarra}, {Santoli}, {Scalise}, {Sciortino}, {Schroeder}, {Seek}, {Sosdian}, {Spandre}, {Speegle}, {Tamagawa}, {Tardiola}, {Tobia}, {Thomas}, {Valerie}, {Vimercati}, {Walden}, {Weddendorf}, {Wedmore}, {Welch}, {Zanetti}, \& {Zanetti}}]{Weisskopf2022}
{Weisskopf}, M.~C., {Soffitta}, P., {Baldini}, L., {et~al.} 2022, Journal of Astronomical Telescopes, Instruments, and Systems, 8, 026002, \dodoi{10.1117/1.JATIS.8.2.026002}

\bibitem[{{Wilms} {et~al.}(2000){Wilms}, {Allen}, \& {McCray}}]{2000ApJ...542..914W}
{Wilms}, J., {Allen}, A., \& {McCray}, R. 2000, \apj, 542, 914, \dodoi{10.1086/317016}

\bibitem[{{Wilson} \& {Weiler}(1976)}]{1976A&A....53...89W}
{Wilson}, A.~S., \& {Weiler}, K.~W. 1976, \aap, 53, 89

\bibitem[{{Xie} {et~al.}(2022){Xie}, {Di Marco}, {La Monaca}, {Liu}, {Muleri}, {Bucciantini}, {Romani}, {Costa}, {Rankin}, {Soffitta}, {Bachetti}, {Di Lalla}, {Fabiani}, {Ferrazzoli}, {Gunji}, {Latronico}, {Negro}, {Omodei}, {Pilia}, {Trois}, {Watanabe}, {Agudo}, {Antonelli}, {Baldini}, {Baumgartner}, {Bellazzini}, {Bianchi}, {Bongiorno}, {Bonino}, {Brez}, {Capitanio}, {Castellano}, {Cavazzuti}, {Ciprini}, {De Rosa}, {Del Monte}, {Di Gesu}, {Donnarumma}, {Doroshenko}, {Dov{\v{c}}iak}, {Ehlert}, {Enoto}, {Evangelista}, {Garcia}, {Hayashida}, {Heyl}, {Iwakiri}, {Jorstad}, {Karas}, {Kitaguchi}, {Kolodziejczak}, {Krawczynski}, {Liodakis}, {Maldera}, {Manfreda}, {Marin}, {Marinucci}, {Marscher}, {Marshall}, {Massaro}, {Matt}, {Mitsuishi}, {Mizuno}, {Ng}, {O'Dell}, {Oppedisano}, {Papitto}, {Pavlov}, {Peirson}, {Perri}, {Pesce-Rollins}, {Petrucci}, {Possenti}, {Poutanen}, {Puccetti}, {Ramsey}, {Ratheesh}, {Sgr{\'o}}, {Slane}, {Spandre}, {Tamagawa}, {Tavecchio}, {Taverna}, {Tawara}, {Tennant}, {Thomas}, {Tombesi},
  {Tsygankov}, {Turolla}, {Vink}, {Weisskopf}, {Wu}, \& {Zane}}]{Xie22}
{Xie}, F., {Di Marco}, A., {La Monaca}, F., {et~al.} 2022, \nat, 612, 658, \dodoi{10.1038/s41586-022-05476-5}

\bibitem[{{Zajczyk} {et~al.}(2012){Zajczyk}, {Gallant}, {Slane}, {Reynolds}, {Bandiera}, {Gouiff{\`e}s}, {Le Floc'h}, {Comer{\'o}n}, \& {Koch Miramond}}]{2012A&A...542A..12Z}
{Zajczyk}, A., {Gallant}, Y.~A., {Slane}, P., {et~al.} 2012, \aap, 542, A12, \dodoi{10.1051/0004-6361/201117194}

\end{thebibliography}
\bibliographystyle{aasjournal}


\appendix

\section{Spectro-polarimetric analysis}\label{sec:app_3ml}
\begin{figure}[b]
\centering
    \includegraphics[width=0.49\textwidth]{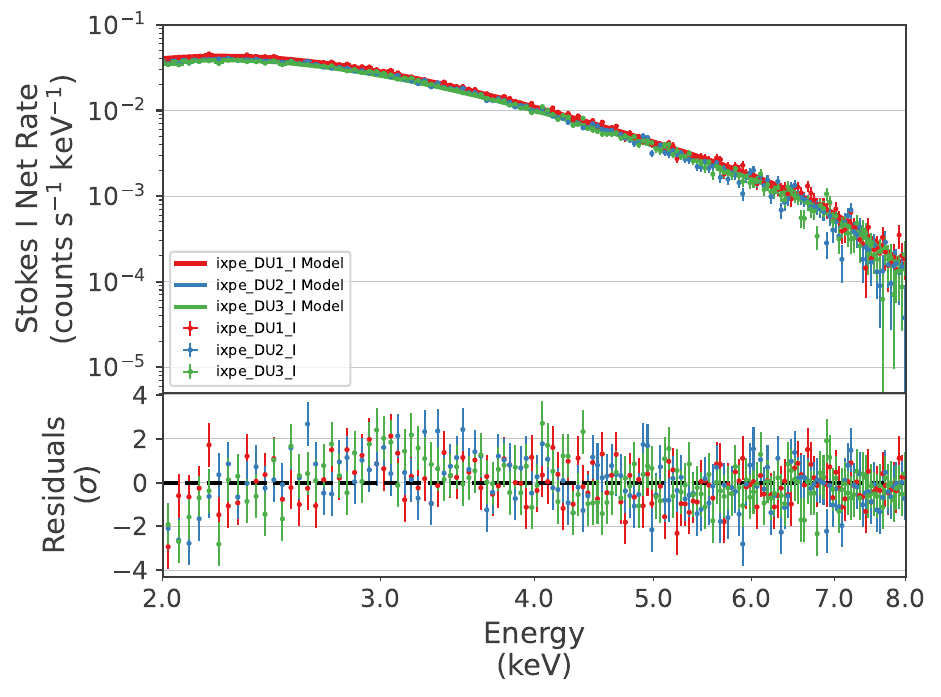}\\
    \includegraphics[width=0.49\textwidth]{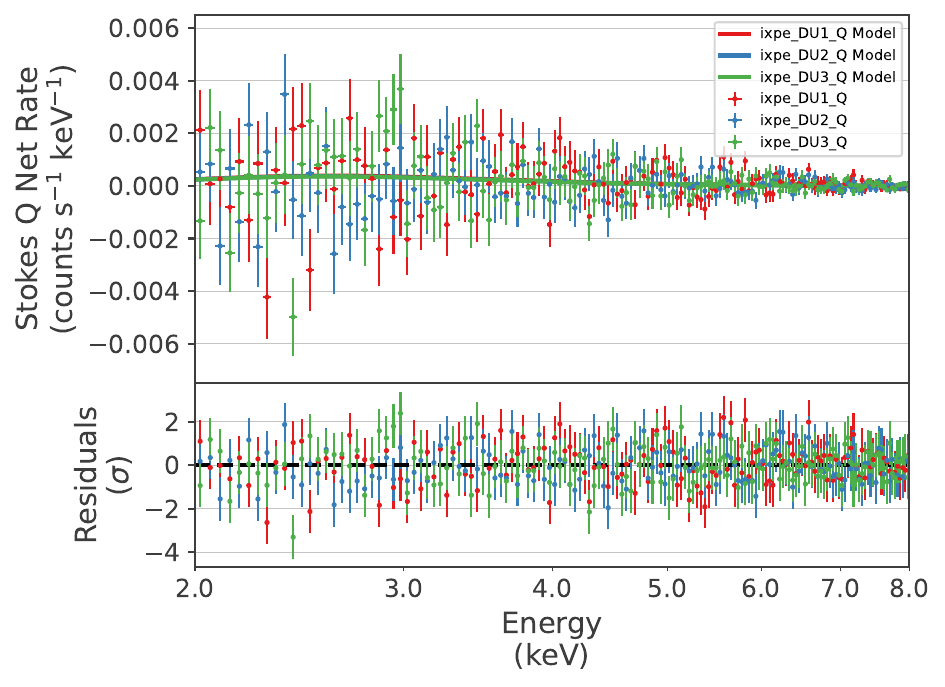}
    \hfill
    \includegraphics[width=0.49\textwidth]{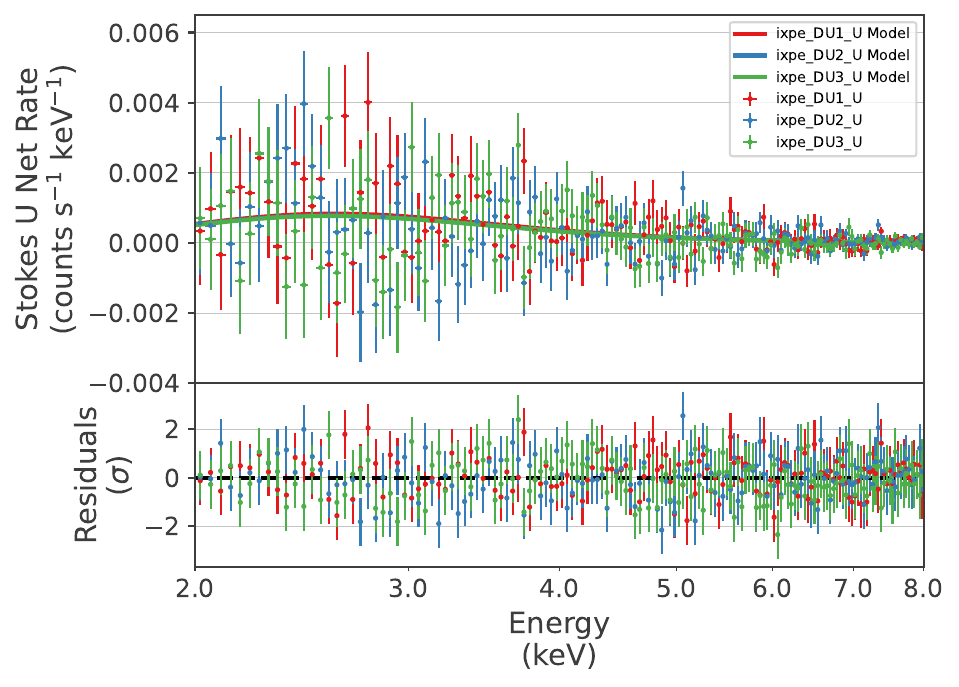}
    \caption{Stokes parameter I (top), Q (bottom left) and U spectra (bottom right) for the three DUs (shown in red, blue and green), along with best-fit models and the residuals from the the \texttt{3ML} analysis (IRFs version \texttt{20240125}).}
    \label{fig:spec_pol}
\end{figure}

In this section of the Appendix, we provide additional plots to support the spectro-polarimetric analysis of the entire PWN, as described in Sect.~\ref{sec:spint}. Fig.~\ref{fig:spec_pol} presents the Stokes I, Q and U spectra for the three DUs, along with their best-fit models and the residuals from the \texttt{3ML} analysis. While the Stokes Q and U spectra are well described by the model, the Stokes I spectral fit shows significant deviations at low energies, most likely due to instrumental calibration issues. These discrepancies become more noticeable as the statistical uncertainties approach the level of systematic errors. To assess the potential impact of these issues on the polarization measurement, we repeated the same analysis in two additional configurations: (i) applying the off-axis vignetting correction to the on-axis IXPE effective area (version \texttt{20240125}, used throughout the paper) using the HEASARC tool \texttt{ixpecalcarf}\footnote{https://heasarc.gsfc.nasa.gov/docs/software/lheasoft/help/ixpecalcarf.py.html}, and (ii) employing the previous time-independent version of the IXPE IRFs (version \texttt{20230526}, released on June 16, 2023). In both cases, while the best-fit power-law index $\Gamma$ varied within the range 1.85--1.98, the resulting PD and PA values remained virtually unchanged. 
%

The IXPE energy response matrix is defined in 275 channels spanning 1–12~keV (0.04~keV per bin), and we analyzed the resulting spectra without additional re-binning, as in other IXPE studies~\citep{2023ApJ...946L..21N}. We verified that rebinning the data to coarser energy grids (e.g., 25 or 100 bins over 2–8~keV) does not affect the spectro-polarimetric results. Given the featureless, non-thermal nature of the source spectrum, we find no need to apply optimized rebinning schemes such as that proposed by~\citet{2016A&A...587A.151K}. Overall, despite the minor calibration uncertainties that affect the spectral analysis, the stability of the polarization measurements confirms the robustness and reliability of our polarimetric results.

\section{Polarization leakage evaluation and correction} \label{sec:app_leak}
\begin{wrapfigure}{r}{0.5\textwidth}
\vspace{-0.7cm}
  \begin{center}
    \includegraphics[trim={0 0.1cm 2cm 1cm},clip,width=0.5\textwidth]{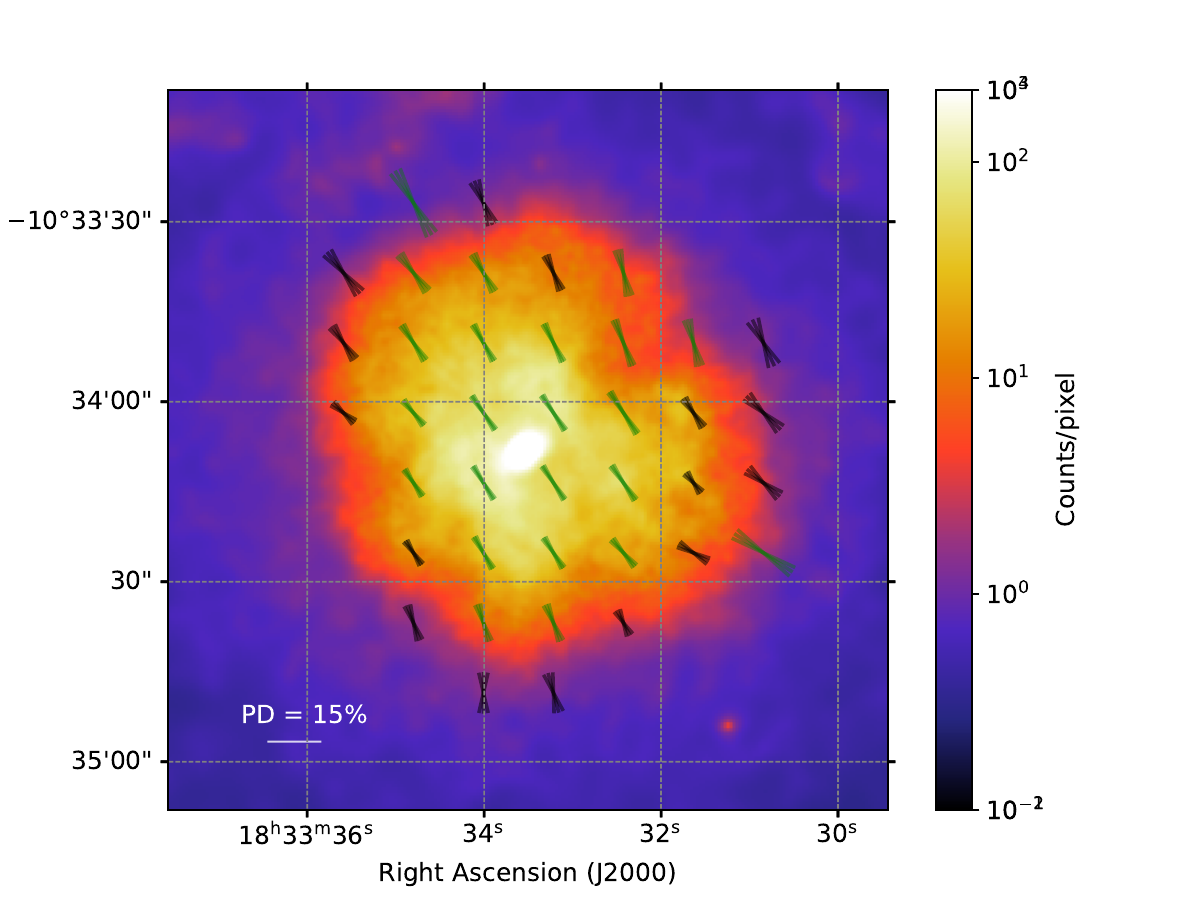}
  \end{center}
  \caption{Polarization map of the PWN as determined from the space-resolved analysis (2--8~keV) in a 0.1944\arcmin\ grid not corrected for the effect of the polarization leakage. As in Fig.~\ref{fig:pmap}, each pixel is correlated with its neighbors and length and orientation of the arrows represent the measured local PD and PA. At the bottom left corner is the reference arrow length for PD = $15\%$. The black and green lines have a $> 2\sigma$ and $> 3\sigma$ significance, respectively. Background image is a \chandra{} image zoomed in over the PWN.}\label{fig:pmap_nosub}
\end{wrapfigure} 

Fig.~\ref{fig:pmap_nosub} shows the uncorrected polarization map of \src{} obtained from the space-resolved analysis (2--8~keV). Compared to Fig.~\ref{fig:pmap}, the outer regions appear to exhibit an outflow pattern, which results from the superposition of the intrinsic, nearly uniform polarization of the source and the radial footprint introduced by polarization leakage.
In this appendix, we provide a detailed comparison of the different techniques used to evaluate and correct the effect of polarization leakage, which is particularly relevant for spatially resolved polarimetric analyses of extended sources such as \src{}. The purpose of this comparison is to validate the robustness of the recovered polarization morphology shown in Fig.~\ref{fig:pmap}.

\begin{figure}[t]
    \centering
    \includegraphics[width=\textwidth]{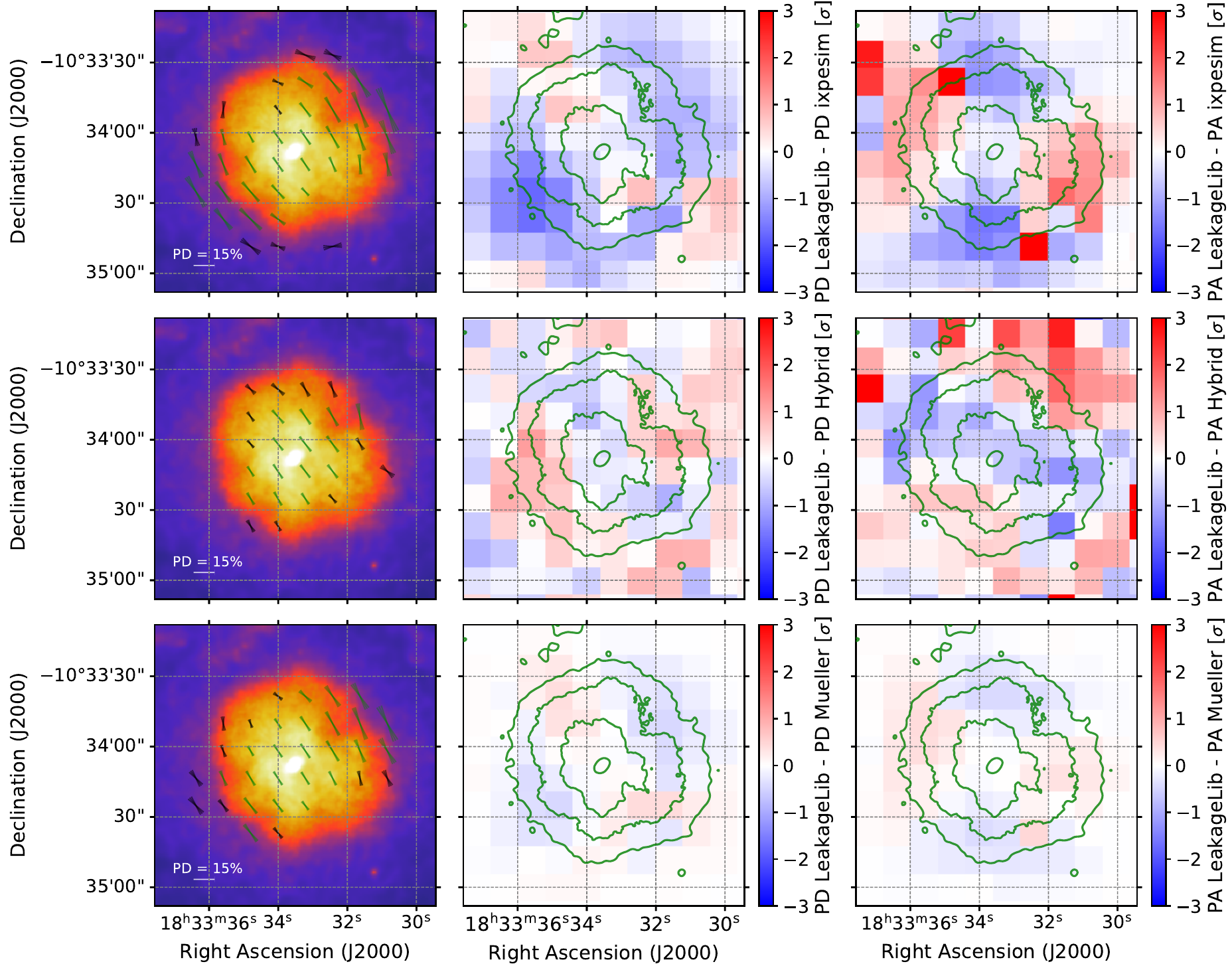}
    \caption{Detailed comparison between the result of the different methods used to estimate and subtract or mitigate the effect of polarization leakage. Going from the top to the bottom, the three rows show the outcome of the \ixpesim/\ixpeobssim\ simulation, the hybrid event reconstruction method and the Mueller matrix approach. The first column shows the corrected polarization maps obtained with each method (2--8~keV). The background image is a Chandra observation zoomed in over the PWN and the polarization measurement is shown using the same convention as Fig.~\ref{fig:pmap_nosub}. The second and third columns display the differences in the PD and PA leakage-corrected maps for each method, relative to the \texttt{LeakageLib} result, rescaled by the sum in quadrature of the corresponding uncertainties. Despite methodological differences, all techniques yield broadly consistent results for both PD and PA across the PWN, confirming the robustness of the polarization analysis.}
    \label{fig:app_leakage}
\end{figure}

As reported in Sect.~\ref{sec:spres}, four independent techniques were employed: the \texttt{LeakageLib} software package, a full \ixpesim/\ixpeobssim\ simulation of the source, a joint machine learning and analytic event reconstruction method, and the Mueller matrix approach, each one described in the sections below. Although each method has some limitations, \texttt{LeakageLib} is currently the most advanced correction technique and was used to generate Fig.~\ref{fig:pmap}. Fig.~\ref{fig:app_leakage} provides a side-by-side comparison of the polarization maps obtained using the three alternative methods relative to \texttt{LeakageLib}, chosen as a reference. The leftmost column of each row shows the corrected polarization map using the indicated method, while the middle and right columns show the residuals in PD and PA compared to \texttt{LeakageLib}, expressed in units of combined uncertainty. The \texttt{LeakageLib} and Mueller matrix methods yield nearly identical outcomes, while the hybrid event reconstruction provides an independent confirmation by directly mitigating leakage effects. The \ixpesim/\ixpeobssim\ simulation slightly overestimates leakage correction but remains within acceptable limits. While small variations exist at the edges of the nebula, where leakage and systematic uncertainties become more prominent, the general structure is stable across methods. Overall, all four techniques yield broadly consistent polarization patterns—especially in the bright, central region of the PWN where the statistical significance is highest—reinforcing confidence in our polarization leakage correction techniques and the robustness of our polarization analysis. 


\subsection{LeakageLib}
The \texttt{LeakageLib}~\citep{Dinsmore2024} package introduces a new model and algorithm for correcting polarization leakage in IXPE data. Compared to the linearized approach of the Mueller matrix formalism, described later in Sect.~\ref{sec:mueller}, this method allows for a more general correlation between the spatial offset of the reconstructed impact point and the inferred polarization direction. Thanks to the implementation of a 2D displacement model, which accounts for both longitudinal and transverse components of the offset, the formalism implemented in \texttt{LeakageLib} can be applied to a more general model of PSF, not necessarily azimuthally symmetric. To better represent the data, this software uses a set of sky-calibrated 2D PSFs, one for each IXPE telescope, derived from a set of bright, weakly polarized point sources observed by IXPE.

For extended sources, the tool provides routines to estimate the expected Stokes maps (I, Q and U) due to polarization leakage based on a \chandra{} observation of the source. The \chandra{} count map is assumed to represent the \textit{true} morphology given its far superior angular resolution compared to IXPE. However, in the current release (v1.1.0), no dependence on the IXPE PSF off-axis angle or event energy has yet been implemented. For \src{}, we modeled the source using the same \chandra{} map employed throughout the paper. The output Stokes maps, containing the predicted signal induced by leakage, are then normalized to match the observed data and properly rebinned using the same grid as in Fig.~\ref{fig:pmap_nosub}. For all IXPE telescopes, the content of the Stokes Q and U maps is finally subtracted from the polarization map obtained in Sect.~\ref{sec:spres}.

\subsection{\ixpesim/\ixpeobssim\ simulation}\label{sec:app_sim}
The \ixpeobssim~\citep{Baldini2022} software allows the user to simulate an IXPE observation starting from a generic model of the source or a \chandra\ observation. However, \ixpeobssim\ alone cannot predict polarization leakage, as this effect is not included in the standard IRFs. To enable leakage prediction, \ixpeobssim\ offers an interface to \ixpesim{}~\citep{dilalla2019}, which implements a detailed photon-by-photon \texttt{GEANT4} simulation of the physical interactions in the GPD. As described in the official documentation\footnote{\url{https://ixpeobssim.readthedocs.io/en/latest/ixpesim.html}}, a list of photons can be simulated in \ixpeobssim\ up to the beryllium window on top of the GPD using the \texttt{xpphotonlist} application. These photons are then propagated throughout the detector using \ixpesim\ and turned into a list of actual photoelectron tracks for the events that trigger the detector. The output file containing the track images (comparable to the IXPE Level-1 data) can be reconstructed using the same event reconstruction algorithm used for the flight data, and a final step using \texttt{xpsimfmt}, part of \ixpeobssim, closes the loop by formatting it in a way virtually identical to the actual IXPE Level-2 data.

Based on a \chandra{} observation of \src{} (Obs ID: 5166), we ran 100 full \ixpesim/\ixpeobssim\ simulations, equivalent to a total exposure time of 40~Ms. We modeled the source as intrinsically unpolarized, so that any pattern detected in the output polarization map could be attributed to the leakage effect. By rescaling and rebinning the resulting Stokes maps with the same binning scheme as in Fig.~\ref{fig:pmap_nosub}, their contribution can be subtracted pixelwise to correct the polarization leakage effect. 
The top row of Fig.~\ref{fig:app_leakage} shows a detailed comparison between the outcome of the \ixpesim/\ixpeobssim\ simulation technique and \texttt{LeakageLib}.
In contrast to all other methods, this approach appears to over-correct the leakage effect, inducing a spurious tangential pattern.
Previous validation studies involving simpler cases, such as point sources, have shown that the \ixpesim/\ixpeobssim\ approach tends to systematically overestimate the level of polarization leakage, likely due to an imperfect tuning of the detector parameters in \ixpesim, given the observed time evolution of the GPD. Nevertheless, the differences measured between the two methods are within 3$\sigma$, although clearly spatially correlated due to the over-prediction of the leakage amplitude by the simulation.

\subsection{Hybrid event reconstruction}
The hybrid event reconstruction is a recently developed method, described in detail in~\citet{2023A&A...674A.107C}, which combines a new machine learning-based prediction of the photon impact point with a standard moment analysis reconstruction of the photoelectron track, commonly used for IXPE data. This joint approach exploits the better performance achieved by convolutional neural networks in predicting the conversion points from the track images, improving the IXPE polarization capabilities (expressed by the modulation factor) and partially mitigating the effect of polarization leakage~\citep{hybrid_2}. 

We reprocessed the IXPE Level-1 data of this observation, publicly available on the HEASARC archive along with the Level-2 data, using the hybrid event reconstruction method. 
This technique is particularly valuable for comparison, as it is the only approach that directly mitigates polarization leakage rather than subtracting it. Thus, it provides an independent and complementary assessment. As described in Sect.~\ref{sec:spres}, we binned the reprocessed data following the same recipe using the \texttt{PMAP} algorithm of \texttt{xpbin} to produce a new polarization map.
The central row of Fig.~\ref{fig:app_leakage} shows the map resulting from this method and its comparison with \texttt{LeakageLib}. Despite the different methodology, the results in the high-significance region of the PWN are fully consistent with \texttt{LeakageLib} within 1$\sigma$.

\subsection{Mueller matrix formalism}\label{sec:mueller}
The Mueller matrix approach, thoroughly discussed in~\citet{Bucciantini2023Leakage}, is essentially a generalization of the PSF for polarized observations, where the elements of the Mueller matrix describe how the intrinsic Stokes parameters mix with each other due to the effect of polarization leakage. The primary assumption of this method is that the displacement of the reconstructed absorption point from the true position is correlated with the direction of the reconstructed polarization plane (preferentially displaced along the direction of the polarization plane). This simplifies the problem to a linear displacement model. The matrix elements can be derived either theoretically from a given PSF, or by fitting the expected functional form using in-flight IXPE data. In either case, computing the Stokes maps of any source is reduced to a simple convolution with a \chandra\ count map. 
For \src{}, we adopted parameters derived from the IXPE observation of Cygnus X-1 (Obs ID: 01002901), one of the brightest, weakly polarized point sources observed by IXPE. Unlike \texttt{LeakageLib}, the tool implementing this formalism assumes a circular, azimuthally symmetric PSF, as the one provided by \ixpeobssim. Although this is an approximation of the real PSF (see \citet{Dinsmore2024} for a detailed comparison), this model still provides a reasonable estimate of the average effect across the three IXPE telescopes. As previously done for \texttt{LeakageLib}, the leakage estimation is based on a \chandra{} observation of \src{}, and the resulting Stokes maps are properly normalized, rebinned and finally subtracted from the uncorrected polarization map, to allow a meaningful comparison. As for the other methods, the subtracted polarization map and its comparison with \texttt{LeakageLib} is reported in Fig.~\ref{fig:app_leakage} (bottom row). The results of these two methods are in excellent agreement, with differences in the PD and PA maps well below 1$\sigma$, most likely due to differences in PSF parameterization.

\end{document}